\documentstyle[pra,aps,epsf,amsmath,amssymb]{revtex}

\newcommand{\bra}[1]{\mathop{\langle\left.{#1}\right|}\nolimits}
\newcommand{\ket}[1]{\mathop{\left|{#1}\right.\rangle}\nolimits}
\newcommand{\braket}[2]{\mathop{\langle{#1}\left.\right|{#2}\rangle}\nolimits}
\newcommand{\avr}[1]{\mathop{\langle{#1}\rangle}\nolimits}
\newcommand{\ketket}[1]{\mathop{\left|\left|{#1}
\right.\right.\rangle\!\rangle}\nolimits}
\newcommand{\brabra}[1]{\mathop{\langle\!\langle\left.\left.{#1}
\right|\right|}\nolimits}

\begin{document}
\draft

\title{Measurement and Physical Content of Quantum Information}
\author{B.\ A.\ Grishanin\thanks{grishan@comsim1.ilc.msu.su} and
V.\ N.\ Zadkov}
\address{International Laser Center and Department of Physics\\
M.\ V.\ Lomonosov Moscow State University, 119899 Moscow, Russia}
\date{September 15, 2002}
\maketitle

\begin{abstract}
Basic quantum information measures involved in the information analysis
of quantum systems are considered. It is shown that the main quantum
information measurement methods depend on whether the corresponding
quantum events are compatible or incompatible. For purely quantum
channels, the coherent and compatible information measures, which are
qualitatively different, can be distinguished. A general information
scheme is proposed for a quantum-physical experiment. In this scheme,
informational optimization of an experimental setup is formulated as a
mathematical problem.
\end{abstract}

\thispagestyle{empty}

\section*{Introduction}\label{section:intro}
Significant progress in quantum physics in the last decade has
essentially influenced the evaluation of the role and qualitative
content of quantum physics, though, certainly, its fundamentals have
been pre-served. In the earlier studies of specific nonclassical
features of quantum systems, an experimenter only had to ensure
suitable macroscopic conditions (via choosing an object, using
macroscopic fields, providing for necessary temperature, etc.). But
today, it has become possible to intentionally vary directly quantum
states of elementary quantum systems. This possibility initiated a
number of novel applied sciences and technologies, such as quantum
cryptography, quantum communications, and quantum calculation physics
[1--4], which exploit nonclassical features of quantum system states. A
comprehensive analysis of quantum specific features characterizing
physical systems (these features are involved in the above
applications) can be found in the latest reviews [5--10] and monographs
[11--14]. Despite the variety of physical mechanisms used for
generation, processing, and transmission of quantum information, all of
them are principally based on the only essential difference between
quantum and classical events. This is non-commutativity of quantum
variables of systems under consideration, which is equivalent to the
nonorthogonality of their quantum states. Because of this circumstance,
it is impossible to consider an arbitrary set of quantum events within
the framework of classical logic, which is the result of the so-called
incompatibility of nonorthogonal states.

Indeed, suppose two quantum states $\ket\alpha$ and $\ket\beta$ are
orthogonal. Then, using them as an algebraic basis for an algebra with
addition in the form of linear subspace union (sp) and product in the
form of intersection, we find that the algebra of quantum events
constructed on $\ket\alpha$ and $\ket\beta$ contains only four
subspaces $\{\varnothing,\ket\alpha,\ket\beta,H\}$, including the empty
 $\varnothing$ and 2D Hilbert $H={\rm sp}(\ket\alpha,\ket\beta)$ spaces.
 It is equivalent to the algebra of
four elements $\{\varnothing,\alpha,\beta,{\cal M}\}$, ${\cal
M}=\alpha\bigcup\beta$ constructed as an aggregate of subsets contained
in the set of two point elements $\alpha$ and $\beta$ (the indices of
quantum states under consideration) with addition of subsets in the
form of their sum and product in the form of their intersection. This
algebra, considered as an elementary example, corresponds to the
classical (bivalent, i.e., Aristotelian [15]) logic underlying
classical physics, where the rule of the excluded third holds: either
$a$ or not-${a}$. In the above example, this rule is expressed as
$$\alpha\cup\beta={\cal M}$$ in terms of indices and as $${\rm sp}
(\ket\alpha,\ket\beta)=H$$ in terms of states. Here, $\ket\beta$ has
the sense of negation of event $\ket\alpha$ and $H$ has the sense of a
known certain event. Note that both elementary events $\ket\alpha$ and
$\ket\beta$ belong to this class of events. If two states are
nonorthogonal, in the relationship ${\rm sp} (\ket\alpha,\ket\beta)=H$
state $\ket\beta$ is not negation of $\ket\alpha$ because it includes a
nonzero projection on $\ket\alpha$ along with the infinite set of other
states $\ket{\gamma}$ existing due to the superposition principle. In
this case, $\ket\alpha$ and $\ket\beta$ are, respectively, eigenvectors
of two non-commuting operators $\hat A$ and $\hat B$ of certain
physical quantities. These quantities cannot be measured simultaneously
since their quantum eigenstates are nonorthogonal. When the above
physical quantities take their possible values, the corresponding sets
of quantum states are incompatible because they cannot be considered
simultaneously within the framework of classical logic.

Below, we present an elementary analysis which reveals a close
relationship between physical quantities and their corresponding
quantum states. Determination of the relationships directly
between coupling states and the quantitative measurement of
information carried by these states is the subject of the
information theory. The purpose of this study is to analyze the
choice of an adequate quantitative measure of quantum information
and its possible role in physics. To this end, we investigate a
number of basic physical models of quantum systems and schemes of
physical experiments in the presence or absence of incompatibility
of quantum states, which is their only fundamental specific
feature causing various quantum effects. In Section I, the main
types of quantum information and corresponding quantum measures
are classified using this criterion. In Sections II and III, we
successively consider coherent and compatible information. The
latter type of quantum information is closely related to
quantifying information efficiency of a given experimental scheme.
This problem is discussed in Section IV.

\section{Incompatibility of quantum events as the basis of nonclassical
specificity and classification of quantum information} The notion of
quantum information formulated simultaneously with the basic laws of
quantum physics is directly related to them and plays a key role in
their interpretation. Any quantum effect (for example, the essentially
microscopic process of the atom spontaneous radiation or a macroscopic
transition to the super-conducting state) can explicitly be related
with the processes of quantum information transformation if this
information is adequately associated with the corresponding ensembles
of quantum states. One can also state that a prototype of quantum
information had appeared before the classical Shannon information
theory was developed. It suffices to recall the Born interpretation of
the physical sense of the wave function or to analyze the information
content of the quantum measurement postulate (the wave function
collapse) [16].

The notion of quantum information was admitted to be important long
ago. However, interest in its application has only recently been
stimulated by the development of modern quantum optics experimental
methods ensuring quantum system control. Progress in this area of
quantum physics enables one to employ quantum information not only as a
useful abstract notion but, also, to manipulate it in actual
experiments in a free manner. Studies of quantum information processing
initiated quantum information physics. This new field of physics is
covered in the literature referred to in the bibliography index [17].

Proceeding from the conventional description of quantum mechanics [16,
18-20], one can suppose that proper physics must deal only with
quantum-physical quantities, whereas quantum states could rather be
studied, irrespective of specific physical variables, within the
framework of mathematics, one of whose fields is sometimes qualified as
the classical information theory. However, a more comprehensive
analysis shows that this is not true: as soon as quantum states are
associated with eigenstates of physical variables characterizing an
actual quantum model, they become carriers of physically meaningful
information. For example, let us consider the mathematical structure of
self-adjoint operators $\hat A$ in Hilbert space $H$, which are used in
quantum-mechanical representations of physical variables. Then, the
spectral decomposition of operator $\hat A=\sum\lambda_n\ket n\bra n$
describes its splitting into mathematical objects of two types: the set
of physically possible values $\lambda_n$ and the set of corresponding
quantum states $\ket n$. The latter objects contain the most general
physical information which is independent of specific values
$\lambda_n$ and characterizes only certain physical events. Each of
these events lies in the fact that a physical quantity takes one of the
values $\lambda_n$.

Information relationships between quantum states are determined by the
dynamic properties of a physical system and, evidently, provide for the
most fundamental description of its dynamic characteristics. These
relationships may characterize the intrinsic dynamics of a quantum
system and its interaction with other systems. Initially, they are
represented as equations for wave functions or quantum state operators.
The theoretical information approach is based on introducing an
adequate quantitative measure of information exchange. In the general
case, being independent of specific physical variables, this measure is
superior to them in the analysis of general dynamic properties of a
quantum system.

If a specific scheme of a physical experiment is not discussed, a
quantum system is described using its information characteristics
which yield quantitative relationships between the quantum states
of a given system and the quantum states of other systems that may
interact with it. For example, the main information content of a
two-level atom radiation is reduced to the fact that the quantum
information carried by this atom is transferred to the
corresponding quantum of the photon field. In this situation, the
quantum receives information on the phase of the initial atom
state, i.e., the information exchange is essentially quantum and
retains coherence of the transformed wave functions (see Section
II). Thus, the information description is more comprehensive than
in the case when it is represented in terms of the atom-field
energy exchange.

The fundamental concept of classical information measure is formulated
in the Shannon information theory [21, 22]. For classical systems of
physical events, a unified quantitative measure of information can be
introduced, which is independent of neither the specific physical
content of this information nor its usage. This measure enables one to
express the asymptotic level of error-free transmitted information
content as the optimized Shannon information content, which is
necessary for combining numerous channels in the presence of noise.

This elegant theory is based on a special property of classical
ensembles that is excluded from the original principles of quantum
physics. This property is reproducibility of classical events:
statistically, it makes no difference whether one and the same
physical system or its information-equivalent copy is at the input
and output. However, the latter situation is impossible in the
quantum world. Obviously, this circumstance initiated the
discussion of the question whether the Shannon information measure
could be applied to quantum systems [23--25].

In this paper, it is shown that the traditional definition of the
Shannon entropy and the corresponding information measure can
successfully be used for analyzing quantum systems also in the case
when fundamental differences between ensembles of classical and quantum
events are properly taken into account.

The quantum theory involves the principle of quantum state
superposition implying the existence of an arbitrary linear combination
$c_1\ket\alpha+c_2\ket\beta$ of states $\ket\alpha$ and $\ket\beta$,
which results in the presence of a continuum set in any quantum system.
This set is Hilbert space $H\ni\psi$ of states, most of which do not
coincide with any orthogonal basis states $\ket n$ associated with a
certain physical quantity described by the operator $\hat A=
\sum\lambda_n\ket n\bra n$. When a quantum system is transformed, the
basis vectors and the entire Hilbert space are also transformed. This
circumstance, which is exploited in algorithms of quantum calculations,
significantly enhances their efficiency due to high concurrency of
operations [3, 4, 13]. However, the above continuum set does not
contain an unlimited information content in the conventional classical
sense.

The point is that an arbitrary quantum state $\ket \alpha\in H$ can be
distinguished from another state $\ket\beta$ only when the states are
orthogonal. The probability of random coincidence of two arbitrary
states is determined by the squared absolute value of the scalar
product, so that the 2D probability density of two equiprobable states
has the form
\begin{equation}\label{Pcoinc}
P(d\alpha,d\beta)=|\braket{\alpha}{\beta}|^2\frac{dV_\alpha
dV_\beta}{D},
\end{equation}

\noindent where $dV_{\alpha}$ and $dV_{\beta}$ are small volume
elements on the sphere of wave functions satisfying the normalization
$\int\ket{\alpha} \bra{\alpha} dV_\alpha=\hat I$ and $D$ is the
dimension of space $H$. Within the framework of the Shannon information
theory, one can find effective number $N_\alpha$ of states $\alpha$
distinguishable by variable $\beta$ and vice versa [26]. Probability
distribution describes the information exchange between two information
variables $\alpha$ and $\beta$, which, according to the Shannon theory,
is characterized by effective number $N_\alpha$ of error-free
transmitted messages that are formed of subsets $A_\alpha$ of indices
corresponding to quantum states $\alpha$. This effective number
(calculated per symbol) is attained for infinitely long sequences of
independently transmitted unit symbols. When each $A_\alpha$ is
associated with appropriate states $\alpha$, all of these states will
be distinguished without error (in the above sense), so that $N_\alpha$
is the number of distinguishable states. It is determined by the
corresponding Shannon information content
$$I_{\alpha\beta}=\int\log_2\frac{P(d\alpha,d\beta)}{P(d\alpha)P(d\beta)}
P(d\alpha,d\beta)$$

\noindent using the formula $N_\alpha=2^{I_{\alpha\beta}}$. For
joint probability distribution (1) (see [27]),
$I_{\alpha\beta}=1-1/\ln 4\simeq0.27865$ bit for $D=2$ and
$I_{\alpha\beta}= (1-{\mathbf C})/\ln2\simeq0.60995$ bit
($\mathbf{C}$ is the Euler constant) for $D\to\infty$. Thus, at
any space dimension $D$, $N_\alpha<2$, i.e. quantum uncertainty
reduces the effective number of distinguishable states even below
a value of two corresponding to one bit. This is caused by high
quantum uncertainty in pure states $\psi$, which, on the one hand,
are analogs of deterministic classical states when only their
orthogonal sets are considered and, on the other hand, contain
internal quantum uncertainty. In the latter case, other
nonorthogonal states may coexist and pure states determine the
corresponding probability distributions $P(x)=|\psi(x)|^2$ for
variables $\hat x$ such that $\psi$ is not their eigenfunction.
Particularly, the corresponding entropy of the $N$-partite state
$\Psi_N=\psi\otimes\dots\otimes\psi$ asymptotically approaches
$N\log_2 D$ at $N\to\infty$, i.e., the pure character of the state
virtually does not influence the uncertainty of all of the quantum
states. The mixed state $\hat I\otimes\dots\otimes\hat I/D^N$
exhibits the maximum uncertainty equal to $N\log_2 D$ bit.

Thus, the first specific feature of quantum information due to
incompatibility of quantum states is the impossibility of extracting a
noticeable information content with spaces of large dimension D without
selecting states. Quantum information always must be selected in
transmission channels transforming a considerable information content
in a distinguishable form.

Let us consider a simple example illustrating the qualitative
characteristics of quantum systems due to incompatibility of all
quantum states. Suppose two two-level atoms are in one and the same
state (Fig. 1a). While the operational sense of the expression {\em in
one and the same state} is quite clear, its qualitative sense
essentially does not coincide with that of this expression in the
classical case. When this example is considered from the classical
standpoint, only two basis states ($k =1,2$) of each atom are taken
into account. Then, when describing the statistic of these atom states,
it makes no difference whether they are assumed to correspond to
different atoms or to one and the same atom. The point is that, in a
combined system, only one state has a nonzero probability and the
knowledge of state $k$ of each atom corresponds to the exact knowledge
of a possible state of the other atom. Thus, when only populations are
considered in the classical case, atoms can be equivalent copies of
each other.

In the quantum case, it is impossible to copy all quantum states in a
similar manner. In addition to the above two states in an atom, there
exist other states $\ket\alpha$ with nonzero probabilities
$\bigl|\braket{\alpha} {k}\bigr|^2$ such that $\ket k$ is not the
proper basis for averaged physical quantities. This circumstance is due
to quantum uncertainty, which is always present in an ensemble of
quantum states (Fig. 1b). It is well known that, for a harmonic
oscillator, this uncertainty manifests itself in nonzero energy of
vacuum fluctuations $\hbar\omega/2$. For two-level atoms, it takes the
form of nonzero values $\hat\sigma_x^2=\hat\sigma_y^2= \hat I$, where
Pauli matrices $\hat\sigma_{x,y}$ have the sense of quadrature cosine
and sine components of atom oscillators. In spite of the fact that
these atoms are in one and the same state, their corresponding
eigenstates are different because the above state refers to different
physical systems. Each of these systems contains its own ensemble of
mutually incompatible quantum states, which are described by
nonorthogonal eigenvectors corresponding to different noncommuting
operators of physical variables as shown in Fig. 1b. Indeed,
mean-square residuals $\bigl(\hat\sigma_x^A-\hat\sigma_x^B \bigr)^2$
and $\bigl(\hat\sigma_y^A -\hat\sigma_y^B\bigr)^2$ are nonzero because
their operators do not commute with the operator of difference of
populations $\hat\sigma_z$ which corresponds to strictly nonzero
difference $\hat \sigma_z^A- \hat\sigma_z^B$. This means that the
eigenstates of these quantum variables for the atoms under
consideration do not coincide, i.e., all of their mutually
corresponding quantum states are by no means copies of each other.

This simple example implies the important general conclusion that, in
the presence of incompatibility, i.e., nonorthogonality in ensembles of
quantum states of two different atoms, these ensembles are always
different. Hence, information on the states of a quantum system at a
certain instant which is obtained by means of any other system
considered at the same instant is never complete. Complete quantum
information on all quantum states of a system at a specified instant
can be provided only by this system itself. The complete information
content can appear in another place (or at another instant) when it is
automatically cancelled at the initial position, which occurs, for
example, during teleportation [2]. Quantum information can be
teleported only to a single receiver, which provides for the
possibility of designing absolutely intercept-secure communication
systems based on quantum cryptography.

The above qualitative concept of uniqueness of quantum information can
be justified quantitatively. Let us consider the squared difference of
projectors on the mutually corresponding wave functions of two atoms
considered at one and the same instant. Integrating the operator of
this squared difference over all possible wave functions, we obtain the
expression yielding a strictly positive value:
\begin{equation}\label{epsilon}
\hat\varepsilon=\int \Bigl(\ket{\alpha}\bra{\alpha}\otimes\hat I_B-\hat
I_A\otimes\ket{\alpha} \bra{\alpha}\Bigr)^2\frac{dV_\alpha}{D}
=\ketket0\brabra0+\frac{1}{3}
\sum\limits_{k=1}^3\ketket{k}\brabra{k}\geqq\frac{1}{3},
\end{equation}

\noindent where integration is performed over the Bloch sphere
including states $\ket\alpha$ with index $\alpha=(\varphi,\vartheta)$,
$dV_\alpha=\sin \vartheta\,d\vartheta d\varphi/(2\pi)$ is a small
volume element, and $V_\alpha{=}D{=}2$ is the total volume of
integration. This expression is similar to the corresponding classical
formula for the rms discrepancy
$$\varepsilon=\sum\limits_\xi(\delta_{\xi_A\xi}-\delta_{\xi_B\xi})^2$$

\noindent between classical indicators $\delta_{\xi_A\xi}$ and
$\delta_{\xi_A\xi}$ of events $\xi_A=\xi$ and $\xi_B=\xi$ related with
random variables $\xi_A$ and $\xi_B$. For any joint probability
distribution $P(\xi_A,\xi_B)=P(\xi_A) \delta_{\xi_A\xi_B}$ that
describes random variables coinciding everywhere, the mean of random
function $\varepsilon$ is zero, i.e., for these probability
distributions, all of the events $\xi_A=\xi$ and $\xi_B=\xi$ are
realized simultaneously and the above quantities are copies of each
other. Being an exact quantum analog of the discrepancy between two
ensembles of classical events, bipartite operator (2) has two proper
subspaces which are formed of singlet and triplet Bell states
$\ket{\ket{k}}$ and correspond to the eigenvalues $\varepsilon_k=1,1/3$
of the mean-square residual. The value corresponding to the singlet
state is three times greater than the value corresponding to the triply
degenerate triplet state. Strict positivity of this operator, i.e., the
absence of the zero eigenvalue, means that, for any joint density
matrix, its averaging yields a nonzero result, which determines the rms
discrepancy of all quantum states. Hence, it is impossible to mutually
copy all quantum states of various systems irrespective of their
states.

\begin{figure}[ht]
\begin{center}
\epsfysize=0.3\textwidth\epsfclipon\leavevmode\epsffile{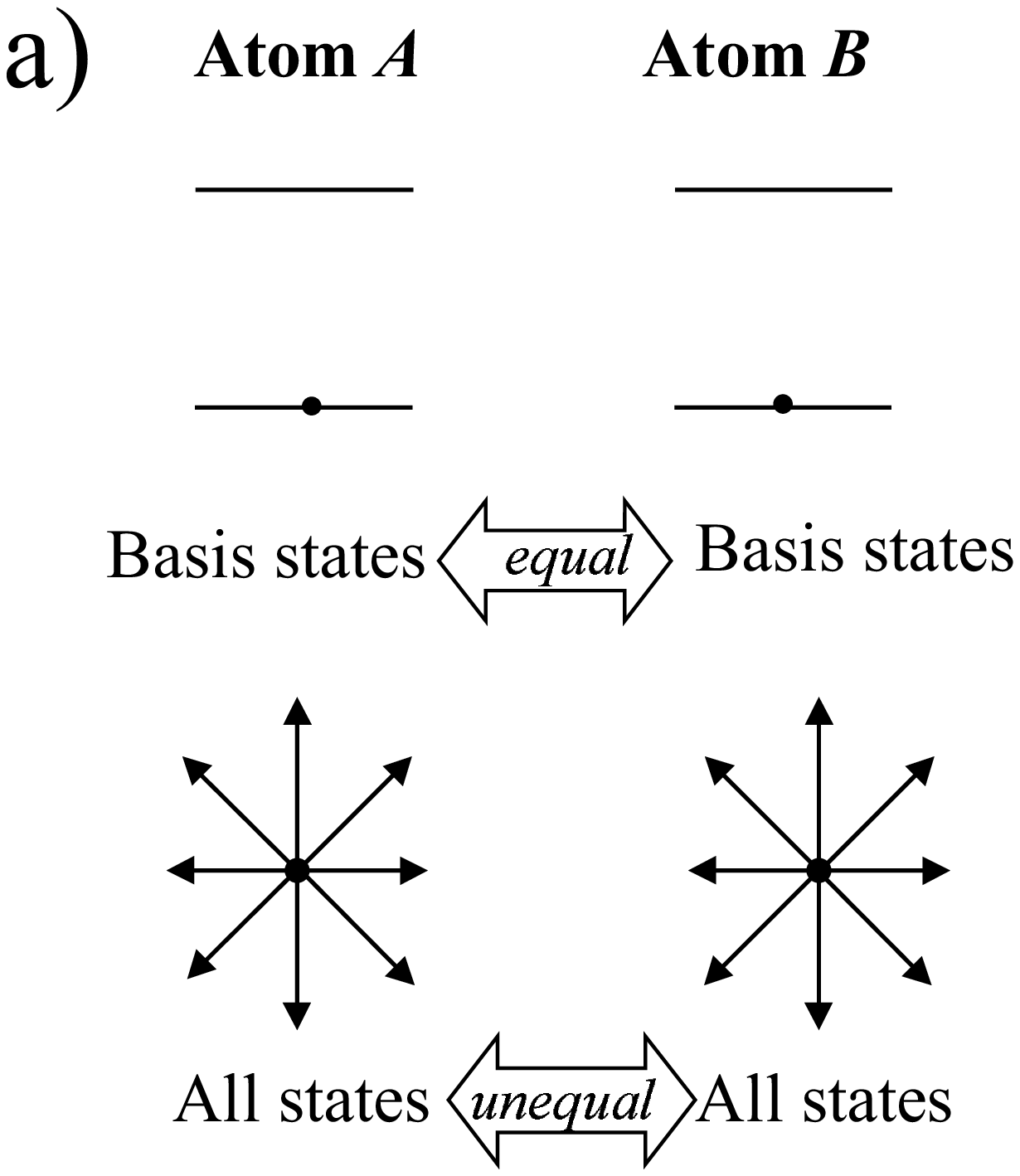}
\hspace{0.04\textwidth}
\epsfysize=0.3\textwidth\epsfclipon\leavevmode\epsffile{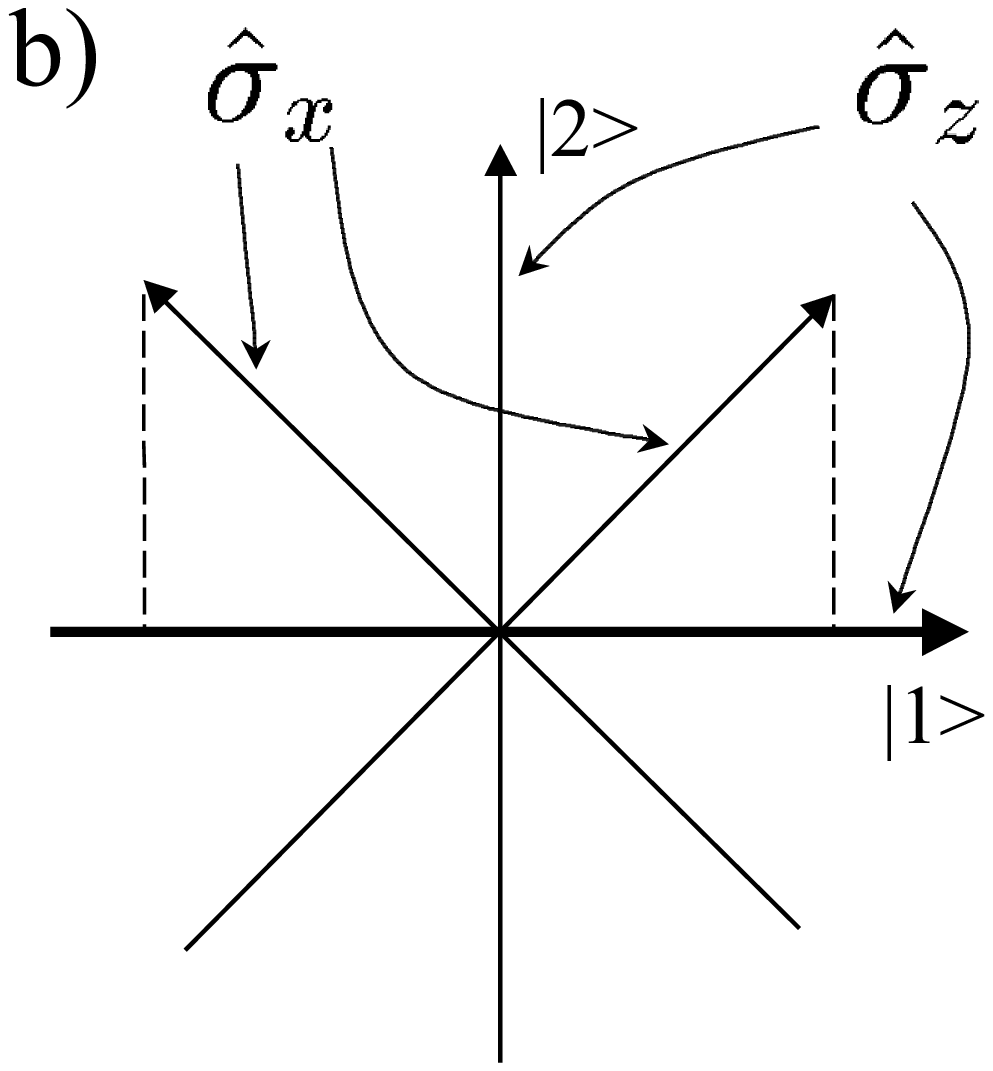}
\end{center}
\vspace{-0.02\textwidth}\caption{Incompatibility of nonorthogonal
quantum states. (a) Equivalence of compatible ensembles of basis states
and nonequivalence of complete quantum state ensembles for two
two-level atoms. (b) Vacuum fluctuations caused by
incompatibility.}\label{fig1}
\end{figure}

The foregoing implies that the key difference between the
classical and quantum types of information depends on whether the
states associated with the information of interest are compatible
or incompatible. The states of different systems considered at one
and the same instant are always compatible. Hence, in the presence
of internally incompatible states, they cannot copy each other,
and the total ensembles of quantum states of one and the same
system which are considered at two different instants are most
often incompatible and, moreover, can copy each other in the
absence of noise retaining the uniqueness of their quantum
fluctuations at every instant. However, the states of two
different systems observed at different instants may be compatible
or incompatible depending on the transformation which couples
these two instants. This factor is rather important for the basic
definitions of quantitative quantum information measures (see
Sections II and III).

Thus, according to the most fundamental classification of quantum
information, which involves the compatibility/incompatibility property
of state ensembles under consideration, the following four types of
information can be recognized: classical information, semiclassical
information, coherent information, and compatible information.

In the case of classical information, all states are compatible and,
within the original form of the Shannon information theory, are
considered as classical on default [21, 22]. However, classical
information can always be transmitted via a quantum channel and is of
interest to quantum physicists as well. A classical channel is
specified by conditional probability distribution $p(y|x)$ of states of
output $y$ at fixed states of input $x$.

For semiclassical information, all input information is specified by
classical states $\lambda$ and output states are characterized by
internal quantum incompatibility as quantum states in Hilbert space
$H$. Nevertheless, output states are automatically compatible with
input states. In the general case, a quantum channel is described by an
ensemble of mixed quantum states $\hat\rho_\lambda$ depending on
classical parameter $\lambda$ [28, 29]. Variables $\lambda$ are
equivalent to input variables $x$ the set of all wave functions
$\psi{\in}H$ is equivalent to output states $y$, and the density matrix
of $\hat\rho_\lambda$ is equivalent to conditional probability
distribution $p(y|x)$ of a classical channel.

In the case of coherent information, the spaces of input and output
states are characterized by internal quantum incompatibility. Being
related by channel superoperator $\cal N$, which transforms the input
density matrix to the output density $\hat\rho_B={\cal N}\hat\rho_A$
[30, 31], these spaces are mutually incompatible. Transformation ${\cal
N}$ determines the flow of quantum incompatible states transmitted from
the channel input to output and is a completely quantum analog of
classical conditional distribution $p(y|x)$, which, in a similar
manner, linearly transforms classical input probability distribution
$p(x)$ to output distribution $p(y)$.

In the case of compatible information, the input and output contain
internally incompatible states but are mutually compatible.

While the first three types of information and the coherent information
measure (which was introduced not long ago) are well known [22, 28,
30], compatible information has been introduced in an explicit form
most recently as a special type of information measure [32]. It is
determined for a combined bipartite system with the compatible input
and output characterized by internal quantum incompatibility.

Coherent and compatible types of information com-prise all possible
qualitatively different types of information in completely quantum
channels. Our study of possibilities provided by applying the
information approach to actual experiments shows that only compatible
information is an adequate tool for analyzing the information
efficiency of an abstract scheme of a quantum-physical experiment.

\section{Coherent information}

\subsection{The Physical Sense of Coherent Information}

According to the classification presented in Section I, one
possible quantitative information measure for a completely quantum
channel is coherent information introduced by Schumacher and Lloyd
[30, 31]. It serves as a quantitative measure of incompatible
information content that is transmitted from one state space to
another. One can consider both the case of one and the same state
space and the case of physically different spaces. A trivial case
of coherent information exchange is the dynamic evolution of a
closed system described by unitary operator $U$: $\hat\rho_B=
U\hat\rho_AU^{-1}$. Here, all pure states $\psi$ admissible by
initial density matrix $\hat\rho_A$ are transmitted unchanged and
transmitted coherent information content $I_c$ coincides with the
initial information content. By definition, the latter is measured
with the Von Neumann entropy $S[\hat\rho_B]= S[\hat\rho_A]$, i.e.,
the corresponding information is described by the expression
\begin{equation}\label{Ic0} I_c=-{\rm Tr}\,\hat\rho_A\log\hat\rho_A.
\end{equation}

\noindent However, this definition requires additional reasoning in
terms of the operational sense of the density matrix. In the
self-consistent theory, this matrix is only a result of averaging the
pure state of a combined system over auxiliary variables. Then,
expression (3) must be considered as the entanglement of input system
$A$ and reference system $R$ for properly chosen pure state $\Psi_{AR}$
(such that ${\rm Tr}_R \ket{\Psi_{AR}} \bra{\Psi_{AR}}=\hat\rho_A$) of
combined system $A{+}R$. Thus, coherent information is measured in
terms of mutually compatible states of two different systems $A$ and
$R$. At the same time, information is transferred from input $A$ to
output $B$, the latter differing from $A$ by a unitary transformation
only.

Information channel $\cal N$ with corresponding noisy environment $E$
(Fig. 2a) must be included in the information system, which completes
the description of its general structure [33].

\begin{figure}[ht]
\begin{center}
\epsfxsize=0.35\textwidth\epsfclipon\leavevmode\epsffile{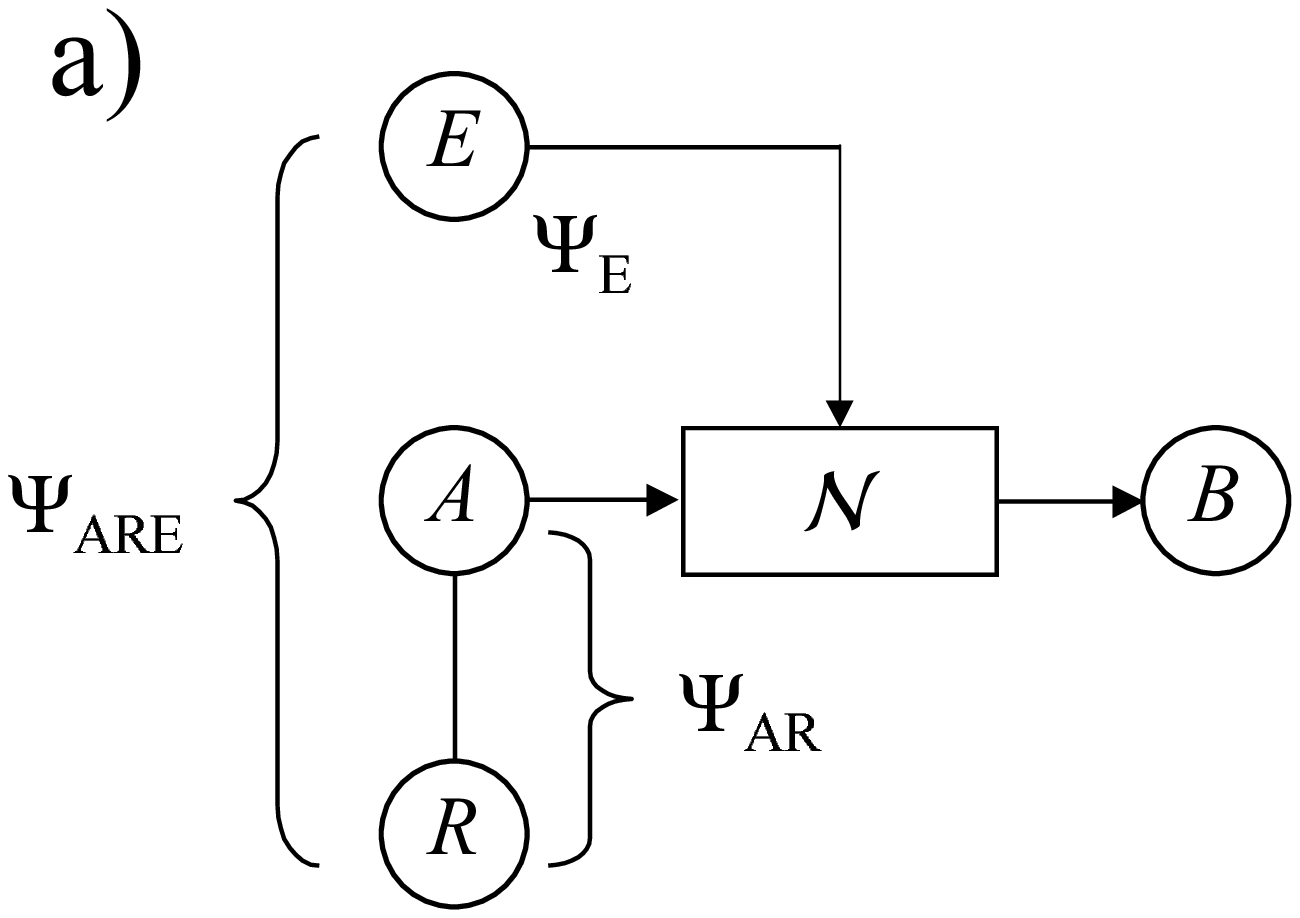}
\hspace{0.02\textwidth}
\epsfxsize=0.45\textwidth\epsfclipon\leavevmode\epsffile{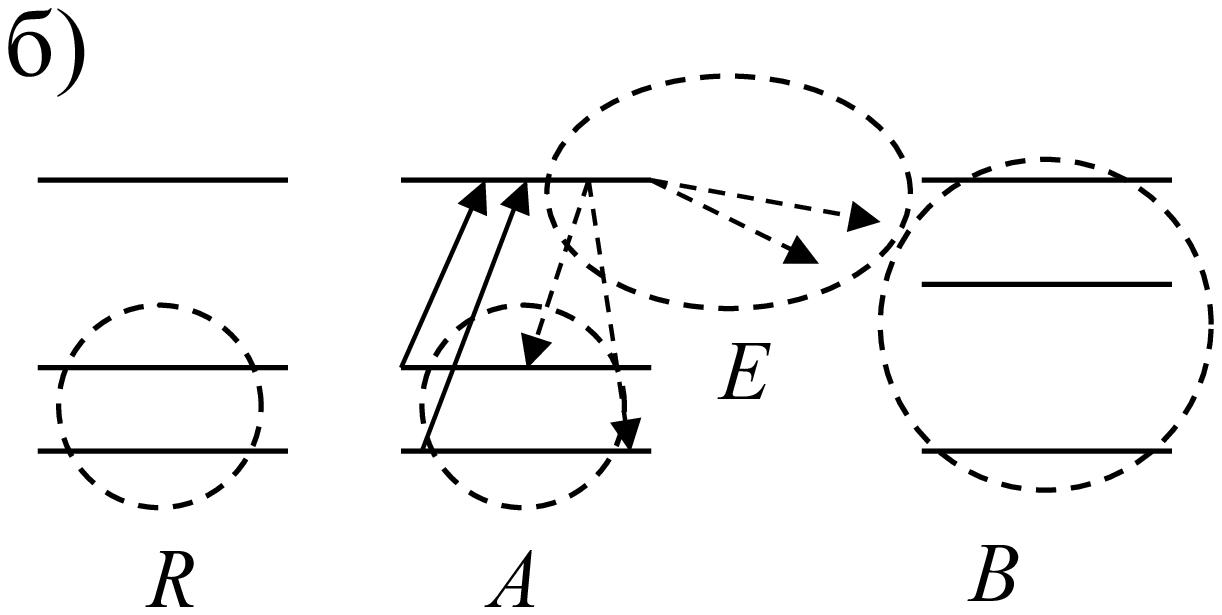}
\end{center}
\vspace{-0.02\textwidth}\caption{(a) Block diagram of a quantum
information system and (b) its possible physical realization: $A$
input, $B$ output, $R$ reference system, and $\cal N$ noisy channel
with noisy environment $E$.}\label{fig2}
\end{figure}

Coherent information for a channel of the general type is defined as
follows [33]:
\begin{equation}\label{Ic}
I_c=S[\hat\rho_B]-S\bigl[({\cal N}{\otimes}{\cal I})\ket{\Psi_{AR}}
\bra{\Psi_{AR}}\bigr],
\end{equation}

\noindent which is in agreement with the Shannon definition of
classical information [34]. Here, $\cal I$ is the unit superoperator
which is applied to the variables of the reference system leaving them
unchanged. The second term in this expression describes the exchange
entropy, which is nonzero only due to the interaction between
subsystems $A{+}R$ and $E$ when ${\cal N}\neq{\cal I}$. Superoperator
$\cal N$ transforms, according to the relationships
\begin{equation}\label{N} \hat\rho_B={\cal N}\hat\rho_A={\rm
Tr}_R\hat\rho_{BR}^{},\quad\hat \rho_{BR}^{}=({\cal N}{\otimes}{\cal
I}) \ket{\Psi_{AR}}\bra{\Psi_{AR}}
\end{equation}

\noindent the states of input $A$ into the states of output $B$, whose
quantum states are compatible with the states of reference system $R$
as before because these systems are not entangled due to the above
transformation. This enables one to consider $B$ and $R$ as
kinematically independent systems described by joint density matrix
$\hat\rho_{BR}^{}$. Taking into account the above circumstance and the
zero, by supposition, entropy of combined system $R\,{+} B\,{+}E$,
relationship (4) can be considered as the measure of entanglement
between output B and combined system $R\,{+}E$ decreased by the entropy
of exchange between the channel $A\to B$ and noisy environment $E$.
Thus, it follows from (4) that, in terms of physical content, coherent
information is a specific measure of preserved entanglement between
compatible systems $R$ and $A$ (remaining after information is
transmitted through the channel $A\to B$) rather than that it directly
serves as a measure of quantum incompatible state flow transmitted from
$A$ to $B$. In the general case, output $B$ may be physically different
from $A$ and even may be described by a Hilbert space of quite
different dimension ($H_B\neq H_A$) [35, 36], which is illustrated by
the physical example of the information system presented in Fig. 2b.

For this system, input $A$ and reference system $R$ correspond to the
ground two-level states of two entangled atom $\Lambda$-systems.
Information channel $\cal N$ is provided by laser excitation of input
system $A$ at the radiation-active upper level. Combined with the
vacuum state, two emitted photons correspond to output $B$, whereas all
of the remaining freedom degrees of the field combined with the excited
atom state form noisy environment $E$.

The elementary carrier of quantum information is a two-level system, an
analog of the classical bit which is conventionally called a
qubit.\footnote{The term {\em qubit} was first introduced by B.
Schumacher [51].} Therefore, it is logical to call the quantitative
measure of specifically quantum coherent information a qubit too.
Obviously, this unit of coherent information corresponds to the use of
a binary logarithm in definition (3), which yields $I_c = 1$ qubit for
a two-level system with density matrix $\hat I/2$ characterizing the
state with the maximum possible quantum entropy. With this state, all
possible quantum states are presented equally and most completely.

Now, let us find out how the quantitative measure of quantum
information can be used in physics. The quantum theory usually is
applied to calculations of certain means of the form $\avr{\hat
A}=\sum\lambda_n\avr{\ket{n}\!\bra{n}}$, where $\lambda_n$ and
$\ket{n}$ are, respectively, eigenvalues and eigenvectors of
operator $\hat A$. This decomposition is the result of averaging
physical quantities represented in terms of probabilities
$P_n^A=\avr{\ket{n}\!\bra{n}}$ of quantum states $\ket{n}$. All
possible variables constitute an infinite set that is much richer
than the set of all quantum states. Therefore, irrespective of
physical values, relationships between physical states provide for
more general information on physical couplings in a more compact
form. The characteristic features of the coherent information
exchange correspond to physical relationships expressed in the
most general form because they are associated with the most
general properties of interaction between two quantum systems
chosen as an input and output and related by a one-to-one
transformation of all possible mixed input states. Actually, the
dependence of coherent information on the parameters of an
information system is of even more fundamental character than
relationships between specific physical quantities. This
dependence is calculated for a number of fundamental models widely
applied in quantum physics [34--36].

As an example, let us consider the well-known Dicke problem [37], where
the information exchange between atoms demonstrates the dynamics of the
same oscillation type [35] as the energy exchange via emitted photons,
which determines radiation damping in a system of two two-level atoms.
In this case, the oscillation dynamics is typical of not only energy
but also the set of other variables. Hence, in order to describe the
general properties of interatomic interaction, it is expedient to
consider coherent information as a measure of pre-served entanglement
rather than the set of various physical variables. The former is a
characteristic of the internal quantum incompatibility exchange between
mutually compatible state sets of the reference and input systems $H_R$
and $H_A$. In the Dicke problem, information exhibits its coherent
character in the time dependence at a large difference between decay
rates of the short-and long-lived Dicke states. The interatomic
exchange coherence is realized by coherent oscillations between both of
these components. Therefore, the lifetime of coherent information is
determined by the short-lived component in contrast to, for example,
the lifetime of the atom population determined by the long-lived state.

Unlike the other types of quantum information, coherent information
enables one to distinguish two qualitatively different information
exchange classes corresponding to the cases when classical information
or quantum state entanglement is used. Only in the latter case,
coherent information is nonzero. Therefore, it is just coherent
information which can be adequately used for finding out to what extent
a quantum information transmission channel retains the capability of
utilizing the output as an input equivalent for realizing situations
where the quantum character of an input signal is of essential
importance. These problems are widely covered in the modern literature
(see [13] and references cited therein). Being a measure of the
entanglement of a quantum system that is preserved by a physical
transformation, coherent information is now of certain practical
interest for quantum information transmission and processing as well as
for the analysis of specific physical models of quantum channels
illustrated below by an example.

\subsection{One-Time Coherent Information}

Proceeding from formal mathematical analogues, we can start
describing a two-sided quantum information channel with the formal
quantum generalization of the Shannon classical mutual information
$I =S_A +S_B - S_{AB}$:
\begin{equation}\label{Istr} I=S[\hat\rho_A]+ S[\hat\rho_B]-
S[\hat \rho_{AB}],
\end{equation}

\noindent This generalization makes sense only when joint density
matrix $\hat\rho_{AB}$ is specified. This matrix is considered as a
direct analog of classical joint probability distribution $P_{AB}$
[38]. Obviously, formula (6) can be applied to quantum systems if we
assume that the states of systems $A$ and $B$ are mutually compatible.
It is {\em a fortiori} true for one-time states of the corresponding
physical systems when they are not overlapped by the parts of one and
the same quantum system containing both an input and output.\footnote{A
similar generalized meaning of coherent information and its calculation
for specific systems are also possible [35]. When systems move at
relativistic velocities, the appropriate relativistic corrections are
necessary. These situations are topical in modern experiments with
entangled quantum states [39] where effects associated with motion of a
measuring system are recorded.} The foregoing justifies the term {\em
one-time} as referred to information that is specified using the joint
density matrix as an initial characteristic. Though one can formally
define quantum information $I$ with expression (6), its physical sense
remains to be seen [40, 41]. This can be attributed to the main
qualitative difference between the classical and quantum information
channels. In general, relationship (5) implies that the quantum input
and output are incompatible as, for example, in the case when the
states of one and the same quantum system are considered at two
different instants. Thus, $A$ and $B$ in (6) cannot be the input and
output that are involved in the definition of coherent information.
Therefore, in the quantum case, in order to apply density matrix
$\hat\rho_{AB}$, one has to physically specify systems $A$ and $B$.
This can be done using the Schumacher definition of coherent
information, which inevitably necessitates modifying expression (6).

\begin{figure}[ht]
\begin{center}
\epsfxsize=0.45\textwidth\epsfclipon\leavevmode\epsffile{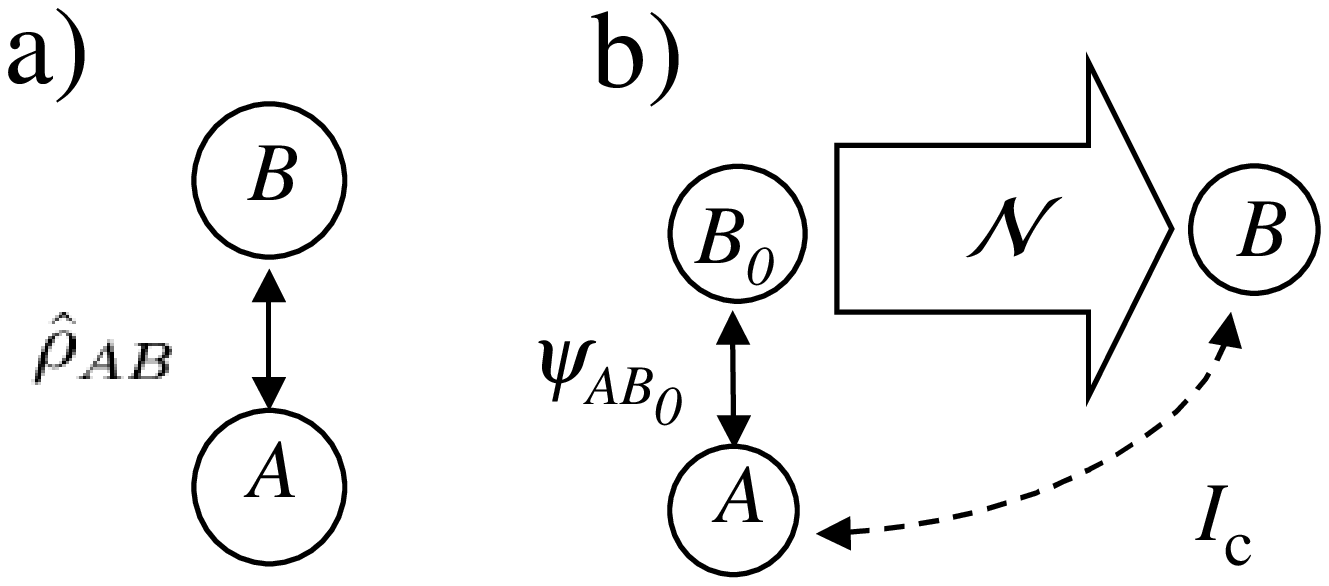}
\end{center}
\vspace{-0.02\textwidth}\caption{Reconstruction of a quantum
information system corresponding to given joint density matrix : (a)
mathematical description of the channel specifying one-time information
and (b) its correspondence with the Schumacher construction
[33].}\label{fig3}
\end{figure}

Adequate physical interpretations of systems $A$ and $B$ are provided
by identifying them, respectively, with the reference system and with
the output of a certain quantum channel associated with specified joint
density matrix $\hat\rho_{AB}$ as shown in Fig. 3. In this case, they
are automatically compatible. Therefore, the actual input of the above
channel corresponds to a certain state of input $B_0$ at an initial
instant rather than to system $A$. At the initial instant, we have
information on A which is not distorted by the channel and is
transformed by $\cal N$ into the final state of output $B$. In this
case, $B_0$ and channel $\cal N$ manifest themselves in the form of
density matrix $\hat\rho_{AB}$ coupling the output and reference system
rather than being explicitly introduced. Specified density matrix
$\hat\rho_{AB}$ properly corresponds to its analog, i.e., density
matrix $\hat\rho_{BR}$ in (5) (which is involved in the Schumacher
method), when the structure consisting of the reference system and
input is characterized by a pure state described by function
$\Psi_{AB_0}$ such that the relationship
\begin{equation}\label{rhoAB}
\hat\rho_{AB}=\bigl({\cal I}{\otimes}{\cal
N}\bigr)\ket{\Psi_{AB_0}}\bra{\Psi_{AB_0}}.
\end{equation}

\noindent holds for a certain channel $\cal N$. Then, the density
matrix
\begin{displaymath}
\hat\rho_A={\rm Tr}_{B_0}\ket{\Psi_{AB_0}}
\bra{\Psi_{AB_0}}
\end{displaymath}

\noindent of the reference system can automatically be represented as
the corresponding partite density matrix ${\rm Tr}_B \hat\rho_{AB}$  of
the system $A\,{+}B$ if one takes into account that the trace with
respect to $B_0$ in (7) is $\cal N$ invariant because this
transformation does not affect subsystem $A$.

The foregoing implies that the corresponding one-time coherent
information can be defined as the Schumacher coherent
information
\begin{equation}\label{I1t}
I_c=S[\hat\rho_B]-S[\hat\rho_{AB}],
\end{equation}

\noindent which, unlike information measure (6), does not contain the
term $S[\hat\rho_A]$. As follows from the above description, in the
general case, the term {\em one-time} does not directly mean that it
refers to the coupling between the system states at one and the same
instant. Actually, reference system $A$ and output $B$ may be
considered at different instants, and the only important condition is
their compatibility. Thus, one-time information couples two compatible
systems all of whose variables are mutually compatible, i.e., described
by commuting operators, in contrast to coherent information, which, in
the general case, couples the incompatible input and output.

Representing coherent information in another form, one-time information
(8) (in contrast to information (6)) is not symmetrical with respect to
$A$ and $B$. Moreover, coherent information may be negative. The latter
circumstance is evident for density matrices $\hat\rho_{AB}$
corresponding to a purely classical exchange between basis sets of
orthogonal states: $\hat\rho_{AB}= \sum P_{ij}\ket{i}\ket{j}
\bra{j}\bra{i}$. Then, all entropies are reduced to classical ones:
$S[\hat\rho_{AB}]{=}S_{AB}{=}-\sum P_{ij}\log P_{ij}$,
$S[\hat\rho_B]{=}S_B{=} -\sum P_j\log P_j$, and $S_{AB}{>}S_B$. A
negative value of coherent information means that the exchange entropy
is so large that it not only makes the preserved information content
vanish but also exceeds the corresponding critical value. In the latter
case, one can assume that $I_c =0$.

\subsection{The Rate of Coherent Information Exchange in
$\Lambda$-Systems}

The information system presented in Fig. 2b plays a special role in
modern applications based on nonclassical features of quantum
information, such as quantum cryptography and quantum computations.
Atomic . systems can be used as basic blocks in these applications.
They are promising carriers of elementary quantum information units
(qubits) that enable one to efficiently store quantum information and
freely manipulate it by means of laser radiation [10, 13]. For the
information system presented in Fig. 2b, the use of the second
$\Lambda$-system as a reference system is physically justified because
the entanglement of two corresponding qubits has a clear physical
meaning as initially stored quantum information. Particularly, the
latter can be applied for performing basic logical operations in
quantum computations involving an output system. The radiation quantum
information transmission channel is of interest because, having
converted an initial qubit into the photon field, one can exploit
various possibilities provided by further high-speed transformations.
It is of interest to find out how rapidly information can be
reconstructed after the qubit-photon field channel is used once.

The reader can find detailed computations of coherent information for
this channel in [36]. Figure 4a shows the dependence of coherent
information on time and the laser field action angle for a symmetrical
$\Lambda$-system. These results are obtained for the input qubit in the
form of the state with the maximum entropy $\hat\rho_A=\hat I/2$. This
qubit state corresponds to the coherent information content that is
independent of individual intensities of two resonant laser fields
affecting the $\Lambda$-system.

\begin{figure}[ht]
\begin{center}
\epsfysize=0.3\textwidth\epsfclipon\leavevmode\epsffile{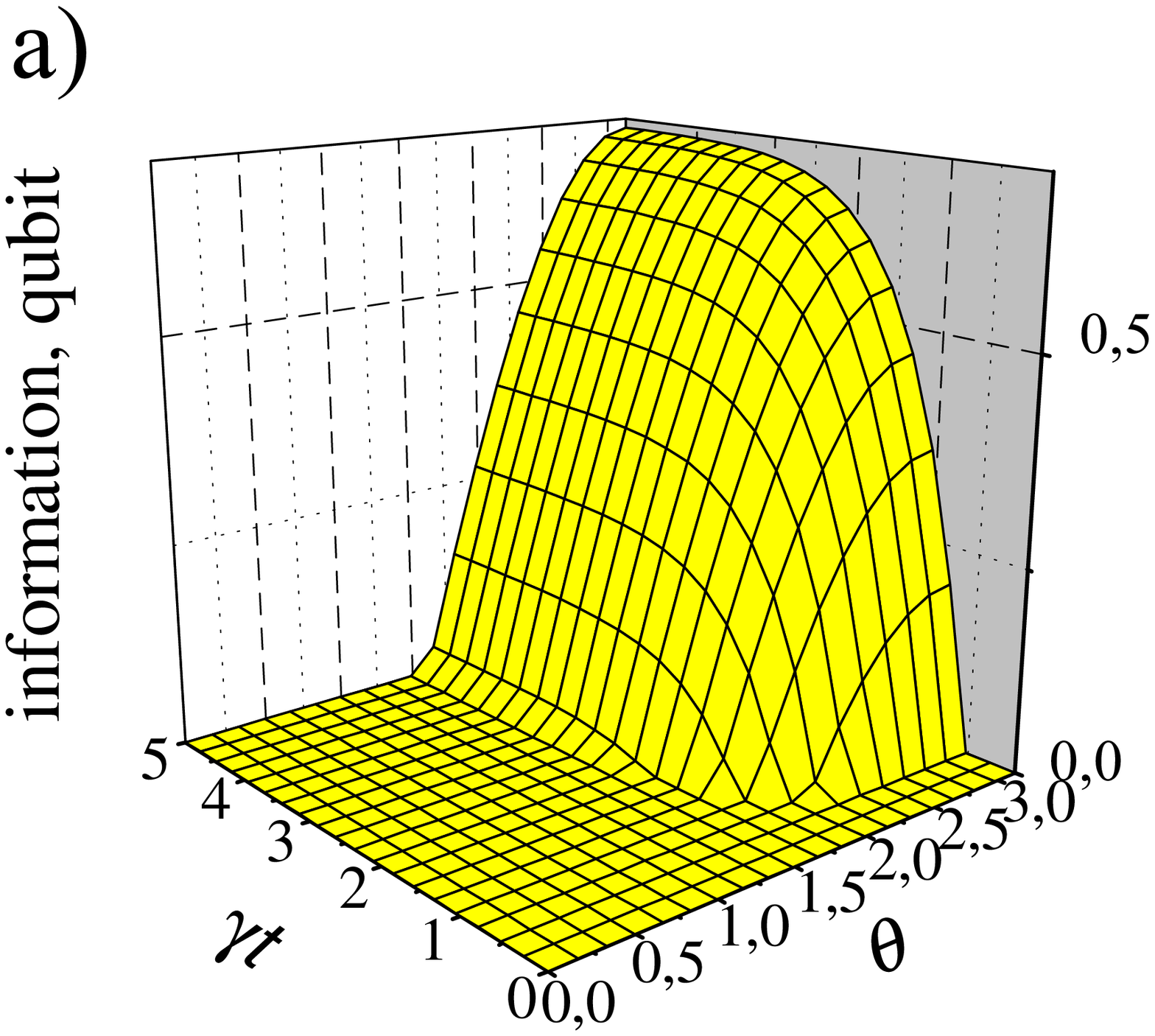}
\hspace{0.02\textwidth}
\epsfysize=0.3\textwidth\epsfclipon\leavevmode\epsffile{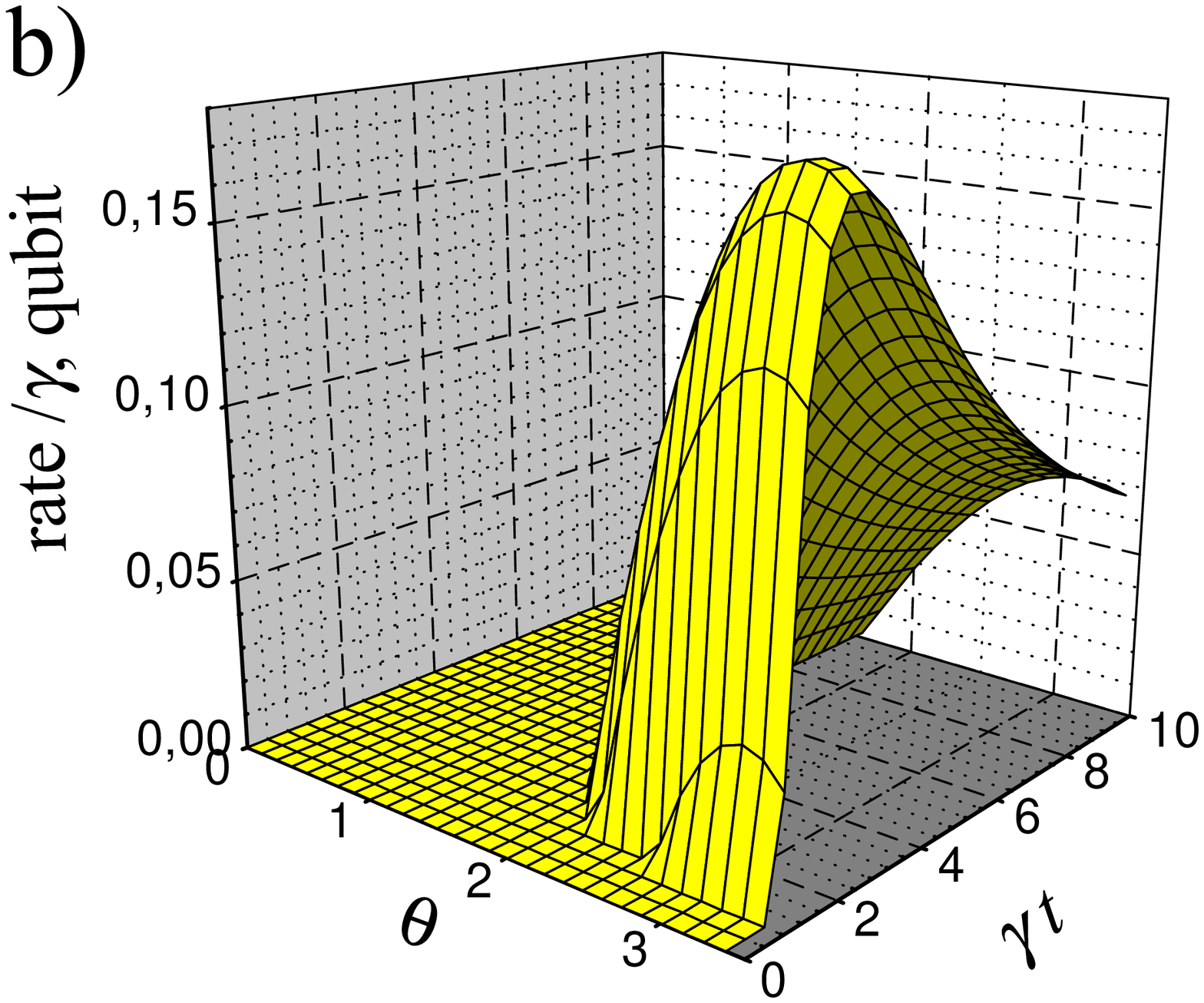}
\end{center}
\vspace{-0.02\textwidth}\caption{(a) Coherent information $I_c$ in a
symmetrical $\Lambda$-system vs. dimensionless time $\gamma t$ and
action angle $\theta=\Omega\tau_p$ for the input state with the maximum
entropy ($\gamma$ is the radiation decay rate, $\Omega$ is the
effective Rabi frequency, and $\tau_p$ is the exciting pulse duration)
[35]. (b) Coherent information rate $R$ vs. cycle duration $t$ and
action angle $\theta$.} \label{fig4}
\end{figure}

One can immediately see from Fig. 4a that there exists the optimum
information exchange level $R=I_c/t$  (where $t=\tau_c$) when the
information channel is used periodically at duration $\tau_c$ of
the exchange cycle, so that the initial state is instantly renewed
after each cycle. Figure 4b demonstrates exchange rate $R$
calculated for a symmetrical $\Lambda$-system with the rates of
radiative decay $\gamma_1{=}\gamma_2{=}\gamma$.\footnote{D.
Bokarev, private communication (2001).} The maximum value reached
by $R$ is $R_0 =0.178\gamma$. Thus, the atom-photon field
information exchange limits the rate of using coherent information
stored in $\Lambda$-systems, i.e., the capacity of the
$\Lambda$-system--photon field coherent information channel. The
order of magnitude of this rate is determined by the rate of
radiative decay of an excited state, whereas its specific value
depends on decay rates $\gamma_{1,2}$ of both radiative
transitions of the $\Lambda$-system. In the limit of a two-level
radiation system, when $\gamma_1=0$ or $\gamma_2=0$, the optimum
rate is equal to $0.316\gamma$.

\section{Compatible information}

For one-time mean values of quantum physical quantities, the
internal quantum incompatibility manifests itself just as
statistical uncertainty, which can be taken into account using the
equivalent classical probability distribution. With the
probabilistic measure
\begin{equation}\label{Pa}
P(d\alpha)=\bra{\alpha}\hat\rho_A\ket{\alpha}dV_\alpha
\end{equation}

\noindent on the set of all quantum states, the mean value of any
operator $\hat A=\sum\lambda_n\ket{n}\bra{n}$ can be represented
as $\avr{\hat A}=\sum\lambda_ndP/dV_{\alpha}(\alpha_n)$, where
$\ket{\alpha_n}= \ket{n}$. Here, $dV_\alpha$ is a small volume
element in the space of physically different states of a
$D$-dimensional Hilbert space $H_A$ ($\int dV_\alpha=D$), which is
represented by the Bloch sphere in the case of qubit, i.e, when
$D=2$ (see Section I). Relationship
\begin{equation}\label{Ea}
\hat E(d\alpha)=\ket{\alpha}\bra{\alpha}dV_\alpha,
\end{equation}

\noindent is the mean of the operator measure (10) which is a special
case of a nonorthogonal decomposition of unit [42], or a positive
operator-valued measure (POVM) [2, 14].

Generalized quantum measurement procedures are described by POVMs.
Unlike a direct measurement in the original system represented by the
orthogonal decomposition of the unit, i.e., by the ortho-projective
measure in $H_A$, a generalized quantum measurement is performed in the
compound space $H_A{\otimes} H_a$. Ha with the appropriate
complementary state space $H_a$ and joint density matrix
$\hat\rho_A{\otimes}\hat\rho_a$, which contains no other information on
$A$ in addition to that contained in density matrix $\hat\rho_A$.

System $A$ is characterized by uncertainty having the form of quantum
incompatibility of the set comprising all of its quantum states.
Generalized quantum measurement (10) transforms this uncertainty into
classical statistical uncertainty of the quantitatively equivalent set
of compatible events in the system $A{+}a$. With this representation,
the coherent relationships typical of the original quantum system are
transformed into the corresponding classical correlations, which have
no quantum specificity. Therefore, this measurement yields a result
that is not equivalent to the original system and cannot provide for
further quantum transformations. This circumstance is the inevitable
pay for information represented in the classical form allowing its free
use. Nevertheless, initial quantum correlations are taken into account
in the statistic of resultant classical states.

Let us assume that two Hilbert spaces $H_A$ and $H_B$ of corresponding
quantum systems $A$ and $B$ are given and joint density matrix
$\hat\rho_{AB}$ is specified in $H_A{\otimes}H_B$. In particular, $A$
and $B$ may correspond to the subsystems of the compound system $A{+}B$
specified at one and the same instant $t$ and can be considered as the
input and output of an abstract quantum channel in an actual physical
system. The determining property of systems $A$ and $B$ is their
compatibility. Hence, the joint measurement represented by two POVMs as
$\hat E_A\otimes\hat E_B$ introduces no new correlations between the
input and output and can be interpreted as an indicator of information
input-output relationships. The corresponding joint probability
distribution is
\begin{equation}\label{Pinout}
P(d\alpha,d\beta)={\rm Tr}\,\bigl[\hat E_A(d\alpha)\otimes\hat
E_B(d\beta) \bigr]\hat\rho_{AB}.
\end{equation}

\noindent Then, the Shannon information $I=S[P(d\alpha)]+ S[P(d\beta)]-
S[P(d\alpha,d\beta)]$ determines the compatible information content
[32, 43]. The physical sense of compatible information depends on a
specific measurement procedure and characterizes the output quantum
information. This information can be obtained via two POVMs, which (in
the form of classical carriers $\alpha$ and $\beta$) select information
on the quantum state of the input, which is transmitted to the output.
As in the case of one-time information, the reference system of the
Schumacher scheme can serve as input $A$ (see Figs. 2a and 3b). Then,
input-output joint density matrix $\hat\rho_{AB}$ can be expressed in
terms of the input partial density matrix and channel superoperator by
formula (7).

Let us consider the special case when $\alpha$ and $\beta$ index all
quantum states in $H_A$ and $H_B$ according to specific POVMs in forms
(10). In this case, compatible information is distributed over all
quantum states and associated with the total internal quantum
uncertainty of input states, which is automatically taken into account
in probability distribution (9). In particular, information that is
contained in quantum correlations observed in the presence of quantum
entanglement between $A$ and $B$ is also involved in joint probability
distribution (11). In addition, here, compatible information is
characterized by operational invariance [44], i.e., all noncom-muting
physical variables are equally taken into account in this information
measure. The above representation of quantum information in terms of
classical probability distributions can be interpreted as a modified
quantum-mechanical representation in terms of classical physical
variables applied in laser physics and discussed by R.J. Glauber in his
lectures [45, 46].

\subsection{Nonselected Information}

It is natural to define information that corresponds to a generalized
measurement specified in form (10) and comprises all quantum states of
a system as nonselected since all quantum states are equally presented
in it and quantum variables are not selected. The opposite situation
occurs in the case of extremely selected information when orthogonal
POVMs are used, which is typical of elementary cryptographic
information exchange schemes [2]. For example, let us analyze
nonselected information depending on the type of joint density matrix
and its main parameter, which specifies the entanglement degree, for
pure and mixed states.

\begin{figure}[ht]
\begin{center}
\epsfysize=0.3\textwidth\epsfclipon\leavevmode\epsffile{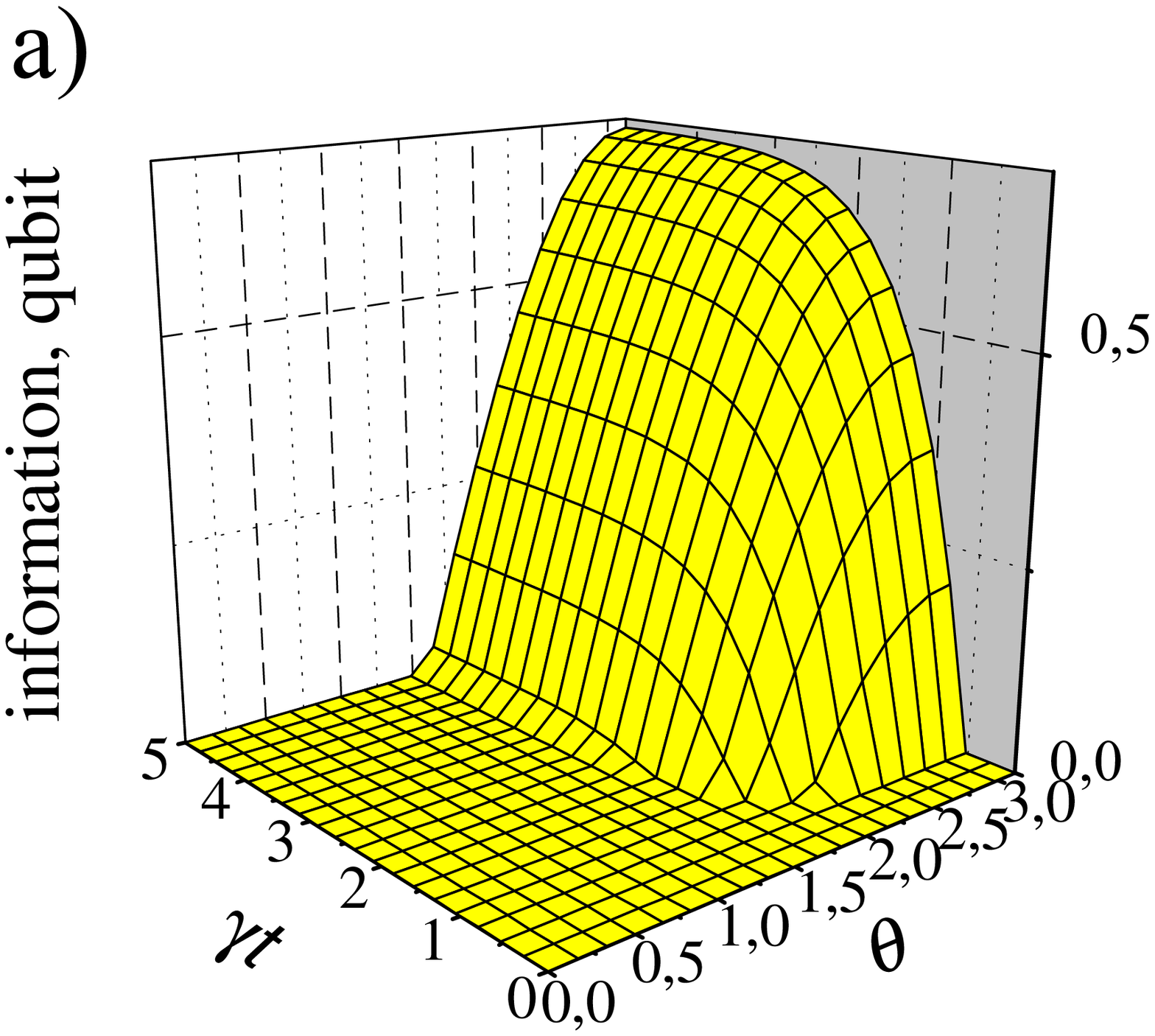}
\hspace{0.02\textwidth}
\epsfysize=0.3\textwidth\epsfclipon\leavevmode\epsffile{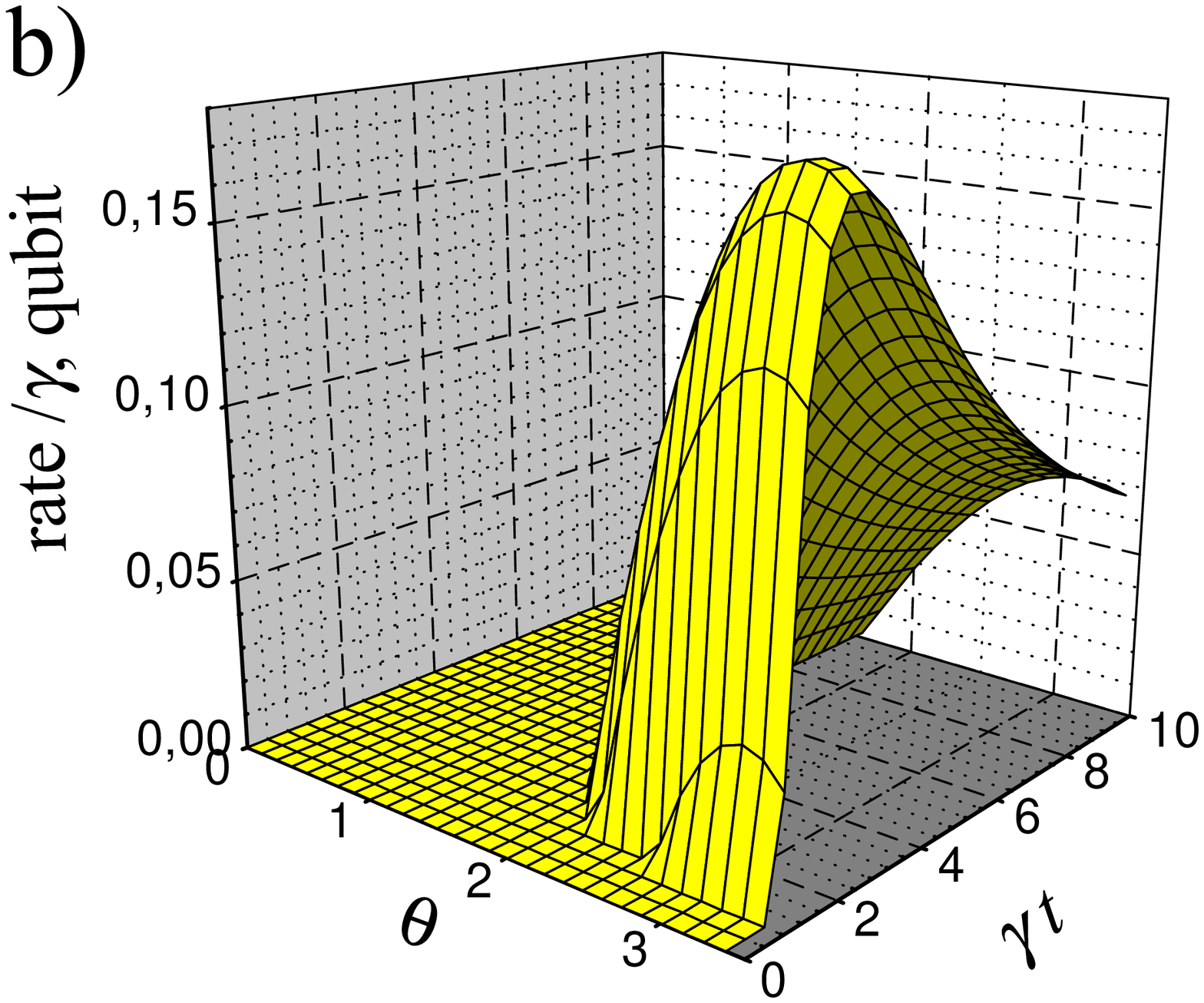}
\end{center}
\vspace{-0.02\textwidth}\caption{Nonselected information $I_u$ vs.
entanglement parameter $q$ for (a) a pure entangled state formed of two
mutually orthogonal basis states with the weights $q/\sqrt2$ and
$q/\sqrt{1-q^2/2}$ and (b) a mixed state formed of a completely
entangled pure state weighted with $q$ and a mixed state which is
weighted with $1-q$ and formed of two equal-weighted pure states
represented as tensor products of orthogonal basis states, so that the
partial input and output entropies are equal to 1 bit for all $q$.}
\label{fig5}
\end{figure}

(i) A pure state is specified by the wave function
\begin{equation}\label{pure}
\hat\rho_{AB}^{(p)}(q)=\ket{\psi_{AB}(q)}\bra{\psi_{AB}(q)}, \quad
\ket{\psi_{AB}(q)}=\sqrt{1-\frac{q^2}{2}}\ket{1}\ket{1}+
\frac{q}{\sqrt2}\ket{2}\ket{2}
\end{equation}
with entanglement parameter $q$. For the limit values $q = 0$ and 1, it
yields a tensor product and completely entangled state, respectively.

(ii) A mixed state is specified by the density matrix
\begin{equation}\label{mixed}
\hat\rho_{AB}^{(m)}(q)=(1-q)\Bigl(\frac{1}{2}\ket{1}\ket{1}\bra{1}\bra{1}+
\frac{1}{2}\ket{2}\ket{2}\bra{2}\bra{2}\Bigr)+
q\ket{\psi_{AB}(1)}\bra{\psi_{AB}(1)},
\end{equation}
where $\psi_{AB}$ is determined in (12). For the limit values $q = 0$
and 1, we obtain, respectively, a mixed state with purely classical
correlations and a pure complete entangled state.

Computations of nonselected information are illustrated by Fig. 5. The
maximum value $I_u =0.27865$ is attained for a completely entangled
state and coincides with the available information content [47]
calculated in [27]. In this context, the term {\em availability} is
considered to mean the possibility of associating with the set of all
possible quantum states of information distinguishable against the
quantum uncertainty background.

\subsection{Selected Information}

Selected information corresponds to generalized measurements with POVMs
$\hat E_A$ and $\hat E_B$ where not all of the quantum states are
equally included. The results presented below are calculated for
selected information in a two-qubit system obtained using measurements
$\hat E_A$ and $\hat E_B$ which combine two different types:
nonselected measurements $\hat E(d\alpha)$ and $\hat E(d\beta)$ and
orthoprojective measurements $\hat E_k$ and $U^{-1}\hat E_lU$
corresponding to direct measurements of orthogonal quantum states.
Then,
\begin{equation}\label{EAB}
\hat{E}_A(\alpha){=}(1{-}\chi)\hat
E_A(d\alpha),\quad\!\hat{E}_A(k){=}\chi\hat E_k,
\quad\!\hat{E}_B(\beta){=}(1{-}\chi)\hat E_B(d\beta),\quad
\!\hat{E}_B(l) {=} \chi U^{-1}\hat E_lU.
\end{equation}
Here, $k,l =1,2$; $\hat E_{k}=\ket{k}\bra{k}$; and $U$ describes
rotation of the second qubit wave function. This rotation is specified
by the transformation $\vartheta$ depending on rotation angle
$U(\vartheta)=\exp(i\hat\sigma_2 \vartheta/2)$ in basis $\ket{k}$ which
is proper for POVM of the first qubit. Discrete results of the output
measurements $(k l)$ complete continual results $\alpha$ and $\beta$,
which corresponds to new variables with an extended spectrum of values:
$a= \alpha,k$ and $b=\beta,l$. In other words, we measure variables
with a combined value spectrum containing a discrete component and a
continuous component. These variables comprise the continuum of all
wave functions and a singled-out orthogonal 2$D$ basis. The limit cases
$\chi=0$ and 1 correspond to nonselected and complete orthoprojective
measurements, respectively. The density matrix has form (13). The joint
probability distribution $$P(da,db)={\rm Tr}\,\hat\rho_{AB}
\bigl[\hat{E}_A(da)\otimes\hat{E}_B(db)\bigr]$$ is represented by the
components $$\begin{array}{ll}P(d\alpha,d\beta){=}(1-\chi)^2
\bra{\alpha}\bra{\beta} \hat\rho_{AB} \ket{\beta}\ket{\alpha}dV_\alpha
dV_\beta,& P(k,l){=}\chi^2\bra{k}\bra{l} \hat\rho_{AB}\ket{l}\ket{k},\\
P(d\alpha,l){=}\chi(1-\chi)\bra{\alpha}\bra{l}\hat\rho_{AB} \ket{l}
\ket{\alpha}dV_\alpha,& P(k,d\beta){=}\chi(1-\chi)\bra{k}\bra{\beta}
\hat\rho_{AB} \ket{\beta}\ket{k}dV_\beta.\end{array}$$ Here, the terms
$P(d\alpha,l)$ and $P(k,d\beta)$ correspond to the information exchange
between discrete and continual measured data for the first and second
qubits. The above relationships imply the following normalization
condition: $$\int\!\!\int P(d\alpha,d\beta) + \int\sum P(d\alpha,l) +
\sum\int P(k,d\beta) + \sum\sum P(k,l) = 1.$$ With (13) and (14), we
have two parameters: the degree of selectivity $0\leqq\chi\leqq1$ of a
combined measurement under consideration, the relative orientation of
orthoprojective measurements $0\leqq\vartheta\leqq\pi/2$ with value
extremes corresponding to the parallel and crossed orientations of the
orthogonal bases of the first and second qubits, and the entanglement
parameter $0\leqq q\leqq1$ (see Figs. 6 and 7).

The plots shown in these figures provide for the following results. The
most unfavorable orientation $\vartheta=\pi/2$ reduces the selected
information content down to zero at $\chi=1$ if the selective
measurement guarantees a nonzero contribution, i.e., if $\chi>0$. At
$\chi>0$, the information content slightly depends on entanglement
parameter $q$. The information maximum $I_s = 1$ bit is reached only
for the degree of selectivity $\chi =1$, i.e., in the case of a direct
measurement.

\begin{figure}[ht]
\begin{center}
\epsfxsize=0.35\textwidth\epsfclipon\leavevmode\epsffile{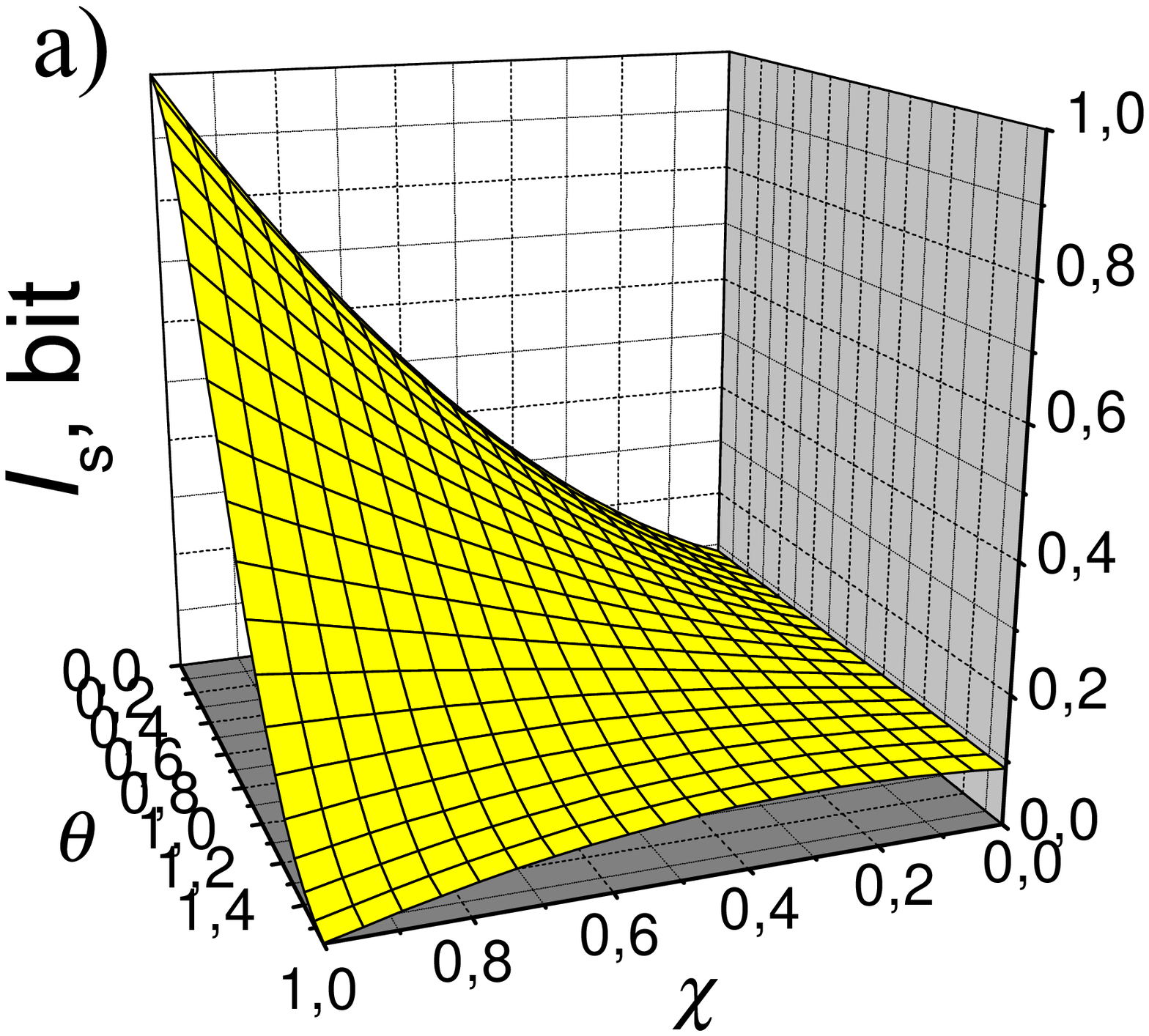}
\hspace{0.05\textwidth}
\epsfxsize=0.35\textwidth\epsfclipon\leavevmode\epsffile{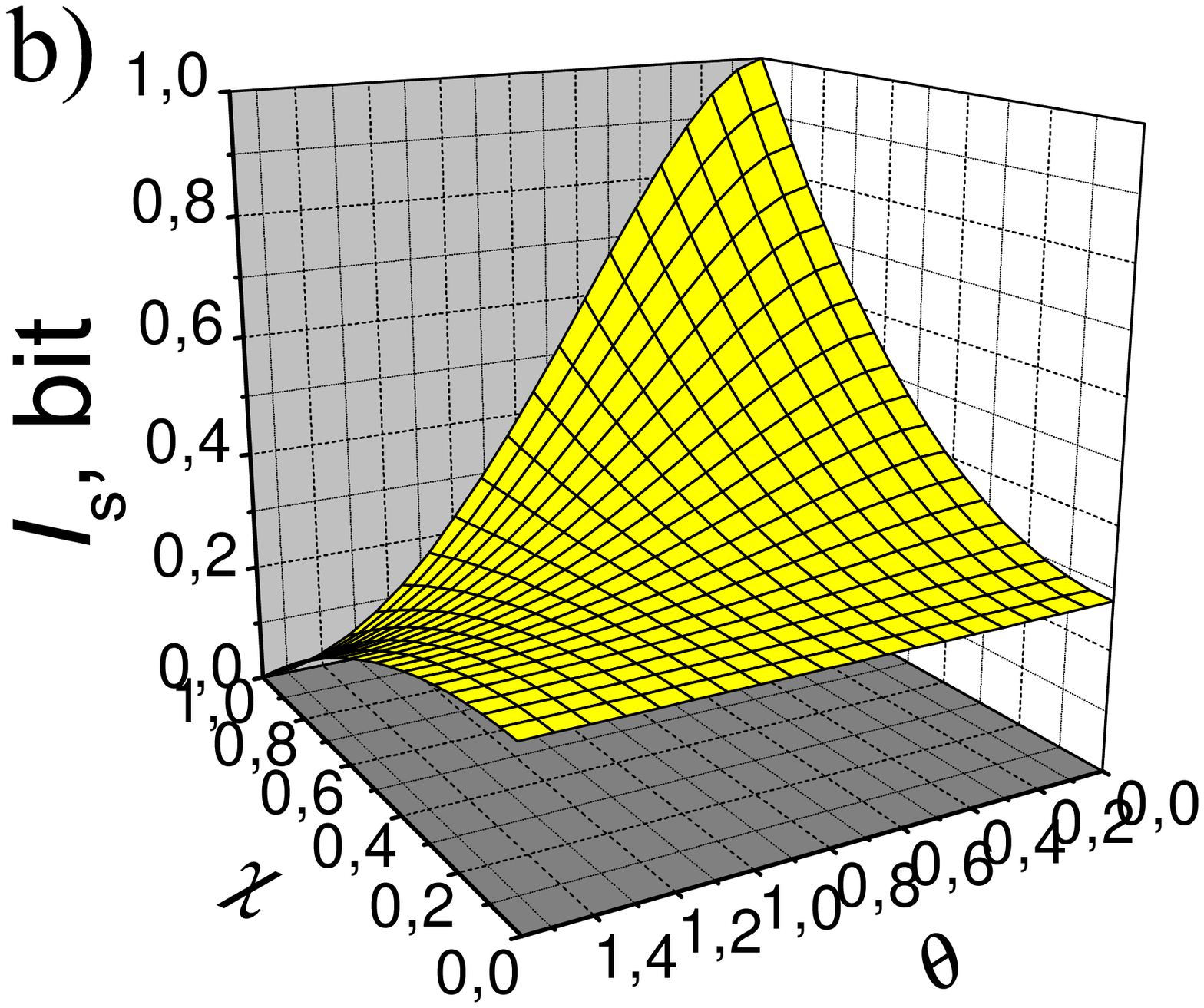}
\end{center}\vspace*{-0.02\textwidth}
\caption{Selected information $I_s$ in a two-qubit system vs. degree of
selectivity $\chi$ and relative orientation of selective measurements
$\vartheta$. (a) in the absence of entanglement for quasi-classical
information communication ($q=0$) and (b) for a pure entangled state
($q=1$).} \label{fig6}
\end{figure}
\begin{figure}[ht]
\begin{center}
\epsfxsize=0.35\textwidth\epsfclipon\leavevmode\epsffile{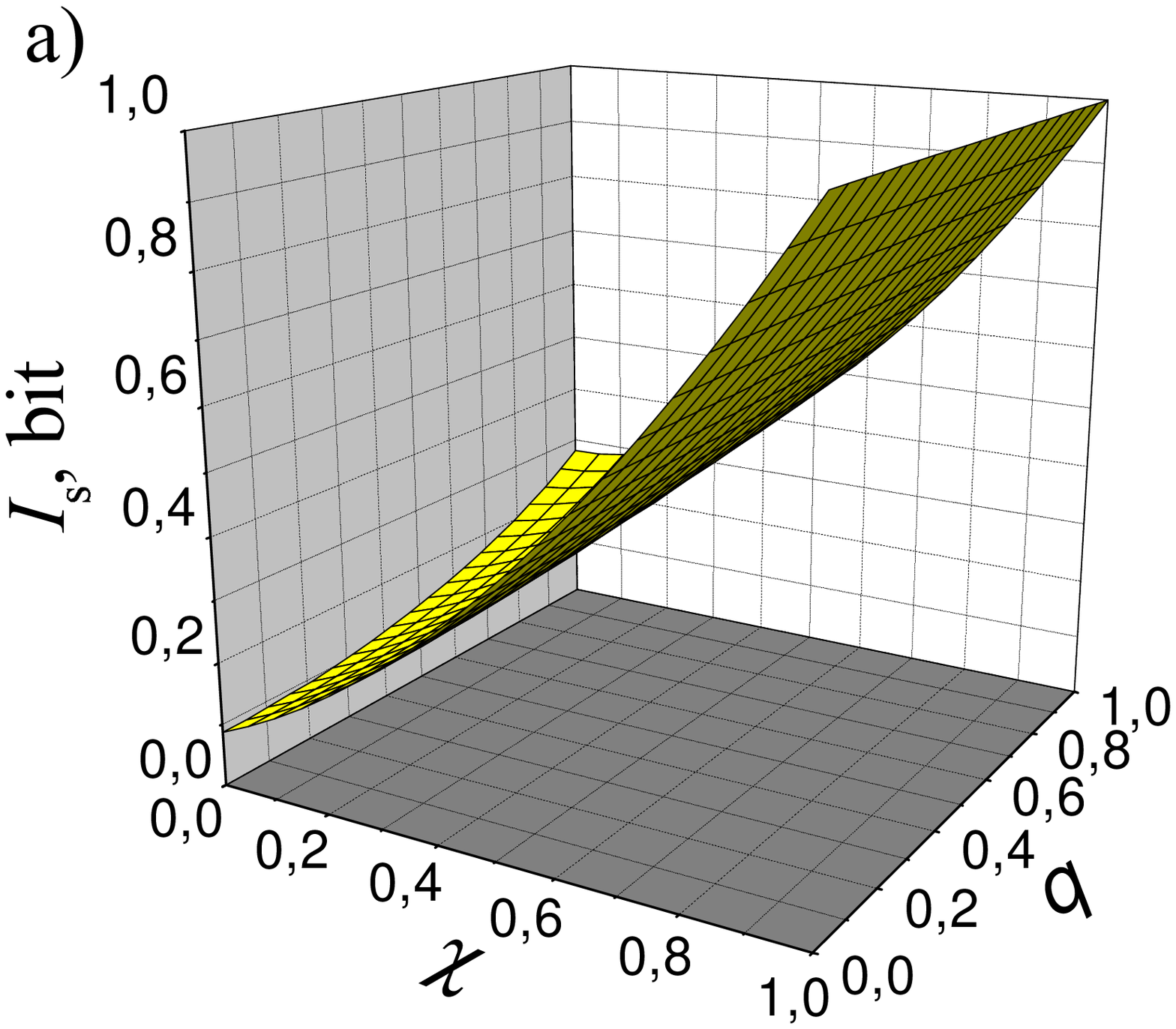}
\hspace{0.05\textwidth}
\epsfxsize=0.35\textwidth\epsfclipon\leavevmode\epsffile{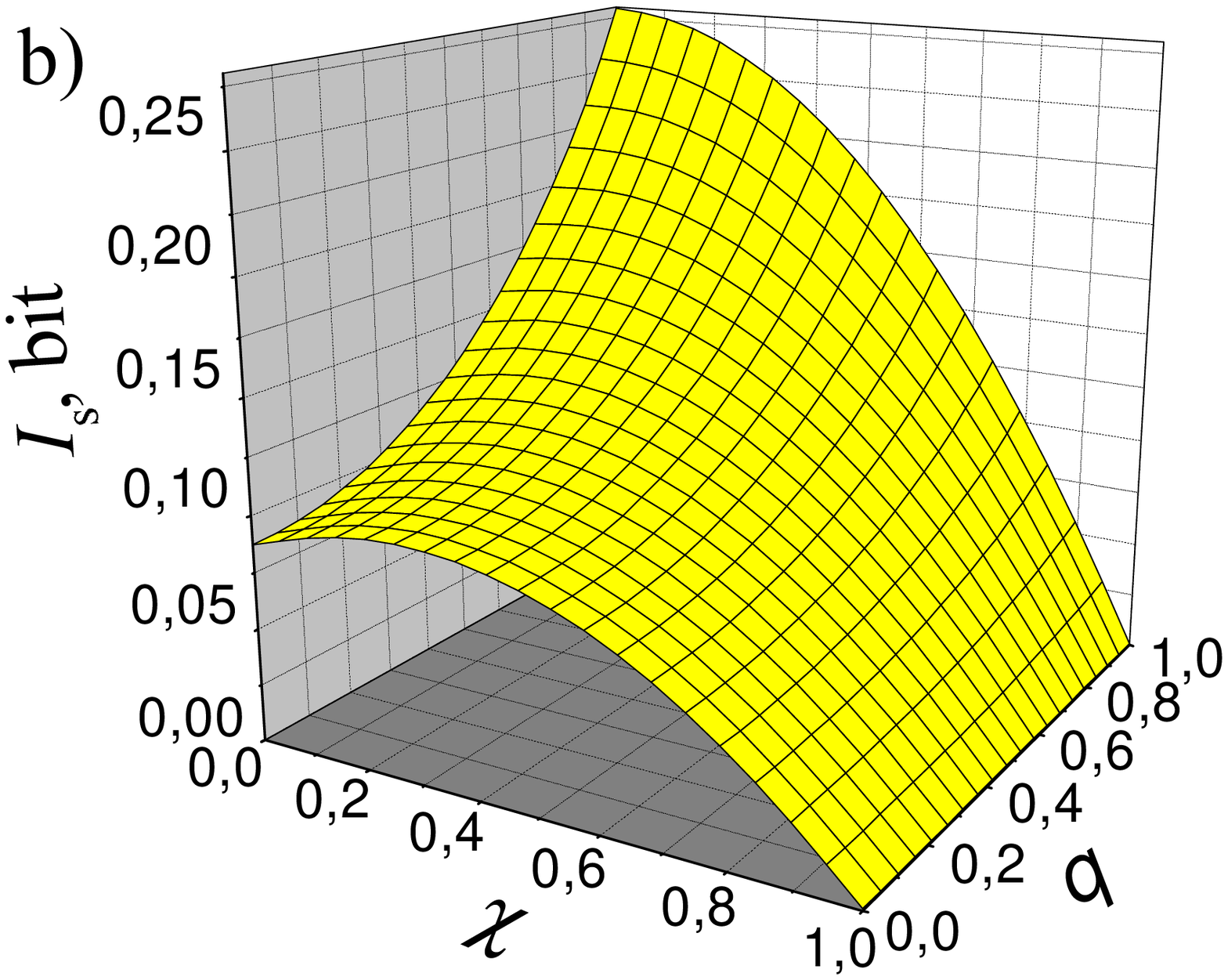}
\end{center}\vspace*{-0.02\textwidth}
\caption{Selected information $I_s$ in a two-qubit system vs. degree of
selectivity $\chi$ and entanglement parameter $q$ for the (a) parallel
($\vartheta=0$) and (b) crossed ($\vartheta=\pi/2$) orientations of
selective measurements.} \label{fig7}
\end{figure}

Note a simple correspondence between nonselected and extremely selected
information, which immediately follows from the physical content of the
corresponding quantum measurement types and holds not only for the
considered 2D example but in the general case as well. Suppose the
completely selected (in our example, $\chi =1$) measurement is
performed at random. This means that we have no {\em a priori}
information about the density matrix structure depending on the input
and output information encoding, i.e., on singled-out orientation
directions of the corresponding spins. Then, the information thus
obtained, evidently, must be averaged over all possible orientations.
The result of this averaging is exactly equal to the nonselected
information con-tent, which is due to the relationship representing the
qualitative content of input-output joint probability distribution (11)
for the above two measuring procedures.

In the case under consideration, when a completely selective
measurement is considered, variables $\alpha$ and $\beta$ in (11)
correspond to discrete indices of basis states $k$ and $l$. With
arbitrary basis wave functions $\ket k$ and $\ket l$ for input $A$ and
output $B$, the corresponding POVMs are $\hat E_k=U_A^{-1}(\alpha)
\ket{k}\bra{k}U_A(\alpha)$ and $\hat E_l=U_B^{-1}(\beta)
\ket{l}\bra{l}U_B(\beta)$, where quantities $U_{A,B}$ describe the
rotation from the initial to measurement basis. Hence, the input-output
joint probability distribution is \begin{equation}\label{PabN}
P_{kl}(\alpha,\beta)=\bra{k}\bra{l}U_A(\alpha)U_B(\beta)\,\hat\rho_{AB}\,
U_B^{-1}(\beta)U_A^{-1}(\alpha)\ket{l}\ket{k},
\end{equation}

\noindent where $\alpha$ and $\beta$ are the parameters of distribution
$P_{kl}$ that determine its dependence on orientations of measuring
procedures. Taking into account POVM (10) and representing the
expression for a nonselective measurement as a sum over projections
with indices $k$ and $l$ of the wave functions $\ket\alpha=
U_A^{-1}(\alpha)\ket0$ and $\ket\beta=U_B^{-1}(\beta)\ket0$, we again
obtain dependences (15) containing $\alpha$ and $\beta$ as information
(in this case) variables. Therefore, distribution (15) over indices $k$
and $l$ simultaneously specifies orientation-angle probability
distribution involved in the nonselective measurement procedure. Hence,
the integrals describing the mean selected information content
exchanged via variables $k$ and $l$ and the integrals describing the
nonselected information exchanged via continuous variables $\alpha$ and
$\beta$ are identical. Thus, a nonselective measurement is equivalent
to the set of completely selective measurements performed
simultaneously for all possible orientation angles of the measurement
basis. The corresponding compatible information automatically takes
into account the uncertainty of the basis orientation at a completely
selective measurement.

\section{Measurement of information available in a physical experiment}

The above analysis based on generalized quantum measurements stimulates
further generalizations that promise a realistic concept of information
content available with a given scheme of a physical experiment, which
can certainly be considered as one of the most important purposes of
the quantum information theory. The main difficulty is to
mathematically describe an information model of a specific experimental
scheme in a sufficiently general form. To this end, first, it is
necessary to mathematically define the notions of input and output,
which is, actually, the most complicated task. Figure 8 demonstrates a
block diagram illustrating the proposed solution.

Being affected by control interactions, the state of an object and its
noisy environment changes. These interactions generate input quantum
information associated either with the dynamic parameters of an object
or with the set of certain quantum states that are of interest. Output
information is measured at the output of the channel described by
superoperator transformation ${\cal N}$. Superoperator measures ${\cal
A}$ and ${\cal B}$ denote transformations realized by control
interactions, and $\hat E_B$ describes the generalized quantum
measurement procedure in the form complying with the POVM.

\begin{figure}[ht]
\begin{center}
\epsfysize=0.3\textwidth\epsfclipon\leavevmode\epsffile{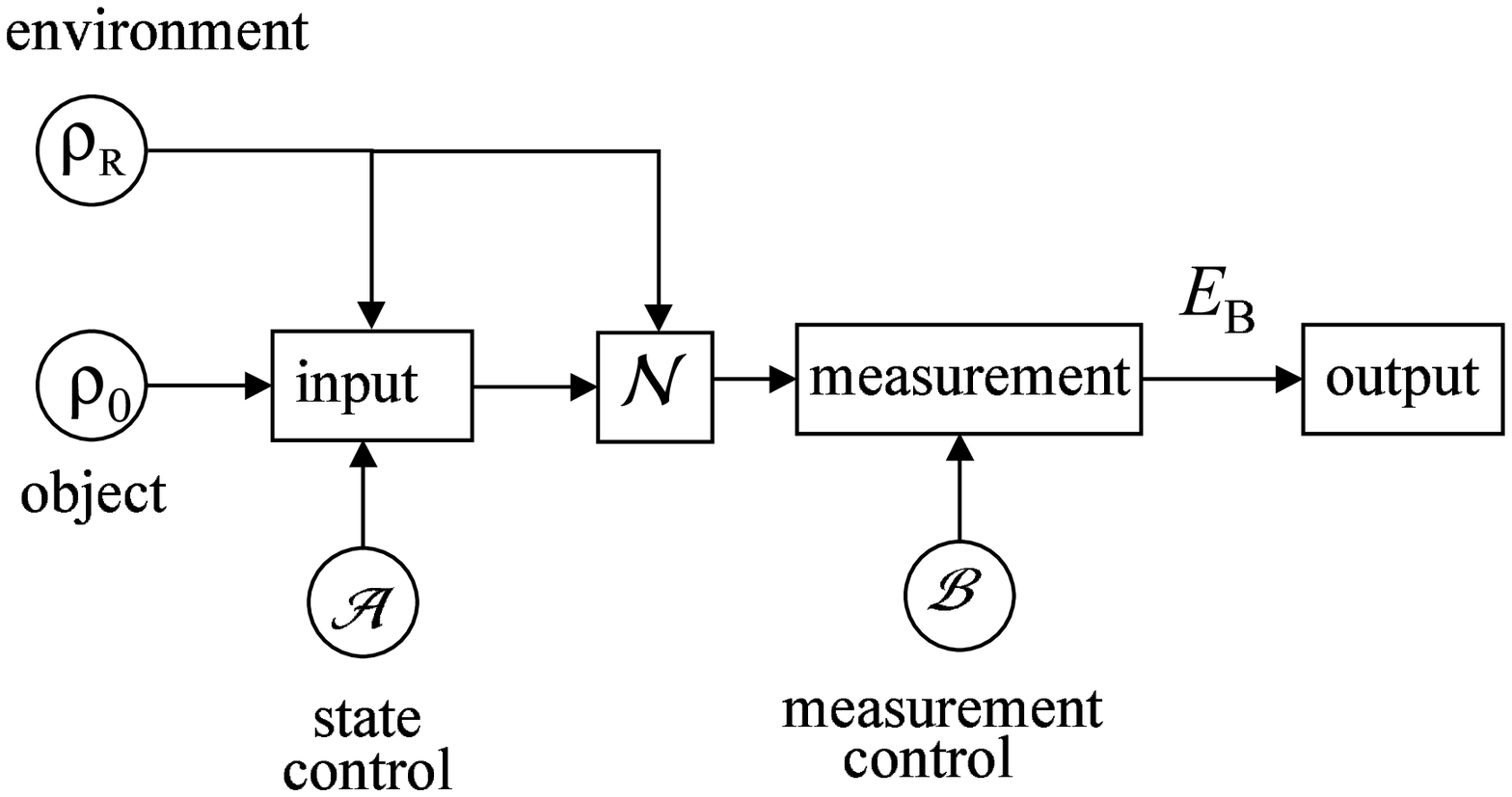}
\end{center}
\vspace{-0.02\textwidth}\caption{Block diagram of an experimental
setup.} \label{fig8}
\end{figure}

This block diagram corresponds to the typical mathematical structure of
the density matrix characterizing a complex system involving two
transformations (${\cal A}$ and ${\cal B}$) which describe the control
and measurement interactions, respectively:
\begin{equation}\label{BNAr}
\hat\rho_{\mbox{\small out}}={\cal B}{\cal N}{\cal
A}\,\hat\rho_{\mbox{\small in}}.
\end{equation}

\noindent Here, $\hat\rho_{\mbox{\small in}}$ and
$\hat\rho_{\mbox{\small out}}$ are the initial and final density
matrices of the set of degrees of freedom essential within the
framework of the mathematical model corresponding to a chosen
experimental scheme. Superoperators ${\cal A}$, $\cal N$ and $\cal B$
describe physical information extraction, transmission to the input,
and measurement, respectively. This Markovian structure of
transformations is not the most general. For simplicity, we assume that
noisy environments corresponding to each transformation are
independent, so that the density matrices can be obtained from
$\hat\rho_{\mbox{in}}$ and taken into account in the structure of
superoperator transformations. Only due to this simplification do we
have the above combination of three superoperators and the input
density matrix and, as a consequence, a relatively simple mathematical
representation of the information structure in terms of the
corresponding decompositions of superoperators $\cal A$ and $\cal B$.
However, in certain cases, it may be necessary to generalize
relationship (16).

Extraction of information always involves using physical interactions
described by the corresponding transformations, which are unitary only
when they con-tain all of the degrees of freedom employed. In addition,
the interaction with the noisy environment must also be included, which
results in non-unitary transformations. Here, we discuss these
transformations for two cases of a possible choice of desired physical
information on a quantum system: (i) dynamic parameters $a$ and (ii)
quantum states $\ket{a}$.

In the first case, physical information is extracted by means of
dynamic excitation of a system which is mathematically described by
unitary operator $U_A(a)$. This operator may depend on control
parameters $c$. Probabilistic measure $\mu(da)$ properly specified must
take into account {\em a priori} information on $a$. Then,
superoperator ${\cal A}$ can be represented as ${\cal A}=\int{\cal
A}_a\mu(da)$ with
\begin{equation}\label{Aa1}
{\cal A}_a=\avr{U_A^{}(a)\odot U_A^{-1}(a)}_E,
\end{equation}

\noindent where the symbol must be replaced by the transformed density
matrix and broken brackets denote averaging over the noisy environment.

In the second case, physical information can eventually be extracted in
a storable form allowing for copying as a result of a certain
generalized measurement corresponding to the set of positive
superoperators
\begin{equation}\label{Aa}
{\cal A}_a=\avr{\ket{a}\bra{a}\odot\ket{a}\bra{a}}_E.
\end{equation}

\noindent In this case, the sum ${\cal A}=\sum{\cal A}_a$ is the
superoperator of a generalized measurement represented using the
averaged standard decomposition $\hat A_i=\hat A_i^+\to\ket{a}\bra{a}$,
which preserves the trace of a completely positive superoperator [48]
with adequately specified operators $\hat A_i=\hat A_i^+\to \ket{a}
\bra{a}$. Since a may also describe continuous variables, one should
use the generalized representation ${\cal A}= \int{\cal A}_a\mu(da)$ in
the form of an integral with measure $\mu(da)$ guaranteeing, with
allowance for idempotency ($\hat P_a^2=\hat P_a$) of orthoprojectors
$\hat P_a=\ket{a}\bra{a}$, that this decomposition corresponds to a
certain POVM: $\int\ket{a}\bra{a}\mu(da)=\hat I$.

In the most general form, the sets of superoperators (17) and (18) are
represented using a certain positive superoperator measure (PSM):
${\cal A}(da)= {\cal A}_a\mu(da)$. This measure is a decomposition of a
completely positive superoperator preserving its trace. The PSM
satisfies the complete positivity (${\cal A}(da)\hat\rho \geqq0$) and
normalization (${\rm Tr}\int\!{\cal A}(da)\hat\rho=1$)conditions. The
latter can be represented in the equivalent form of the unit operator
$\int{\cal A}^*(da)\hat I=\hat I$ preserved under the effect of
Hermitean-conjugated PSM ${\cal A}^*$.

Again, it is of interest to consider the special PSM described by
expression (10) with states specified in Hilbert spaces $H_A$ and
$H_B$ which correspond to transformations ${\cal A}$ and ${\cal
B}$. This PSM associates the information content experimentally
extracted directly from quantum states, which yields the most
explicit description of fundamental constraints due to the quantum
nature of information. In this situation, the output information
is represented in a stable classical form, which potentially
enables numerous users to employ it simultaneously. This
characteristic feature of classical information may initially be
involved on default in the meaning of the term {\em information},
at least, when concerned with experimental physical information
unlike the physical con- tent of coherent information discussed in
Section II.

Applying the above approach to the superoperator of a measuring system
${\cal B}{=}$ $\int{\cal B}(db){=}$ $\int{\cal B}_b\nu(db)$, where
${\cal B}_b$ has form (18), we can represent the input and output
information as classical variables $a$ and $b$ describing the desired
information in both cases (i) and (ii). The joint probability
distribution corresponding to these variables is
\begin{equation}\label{Pab}
P(da,db)={\rm Tr}\,{\cal B}(db){\cal N}{\cal A}(da)\,\hat\rho_{\rm in}.
\end{equation}

Evidently, this distribution is always positive and normalized to
unity. It statistically associates the desired variables and output
information that is extracted using an experimental setup. The
information efficiency of the latter can be expressed in a quantitative
form as the Shannon classical information corresponding to the above
probability distribution. This quantity can be used as an optimality
criterion for optimizing the experiment by available control
parameters.

Note that, for choice (i), states $\ket a$ and $\ket b$ are not
supposed to be mutually compatible and, in the general case, they may
correspond to noncommuting variables. In the trivial limit case, they
may coincide or differ by a unitary transformation, i.e., all quantum
information is sent with the error probability equal to zero. In this
case, the internal quantum uncertainty of this system does not enable
one to use distribution (19), which establishes the one-to-one
correspondence between $a$ and $b$. If the states belong to physically
different sub-systems, they, nevertheless, may contain quantum
correlations due to the corresponding superoperator transformation of
channel ${\cal N}$. The simplest example is the superoperator ${\cal
N}=U_{AB}\odot U_{AB}^{-1}$, which describes a unitary transformation
entangling the input and output states.

Control parameters c may be fixed or chosen from a certain set
$c\in\mathbb{C}$ of necessary values. In the latter case, information
can be optimized according to the above criterion. The presence of
unknown {\em a priori} distribution $\mu(da)$ for dynamic parameters a
in this information structure is not caused by quantum specific
features of the problem, i.e., the problem of a priori uncertainty has
to be treated by the methods employed in the classical theory of
optimal statistical decisions [49]. In the most general case,
transformations ${\cal B}_b$ (see (18)) of a measuring system can be
described with an arbitrary PSM.

Consequently, PSMs ${\cal A}(da)$ and ${\cal B}(db)$ cover a very wide
range of possible types of object quantum state control when the above
quantum measurement procedure is implemented in the experimental
procedure under study.

\section{Conclusion} In this paper, the most general classification of
quantum information is proposed. It is based on compatibility/
incompatibility of the input and output states of a quantum channel.
According to this classification, all possible types of information are
categorized as classical, semiclassical, coherent, and compatible types
of information.

The physical content of coherent information is the amount of
information contained in internally incompatible states that is
exchanged between two systems and quantified as the entanglement
preserved between the output and the reference system. The entanglement
with the latter is used to measure information at the input and output
of a quantum channel specified by a superoperator transformation. Here,
we introduce the concept of one-time coherent information with the
information channel represented by the corresponding joint density
matrix. With this concept, two approaches to determining quantum
information are adequately associated with each other. According to the
first approach, the channel is specified by a superoperator
transformation of the input density matrix, and according to the
second, by the joint input--output density matrix. The coherent
information exchange rate calculated for the channel between a
$\Lambda$-system and the field of free photons yields the upper bound
equal to $0.178\gamma$ for a symmetric $\Lambda$-system and
$0.316\gamma$ in the absence of this constraint.

The necessity of introducing compatible information as an adequate
characteristics of quantum information exchange between two compatible
systems of quantum states is justified. Compatible information is
expressed in terms of the classical information theory despite the
presence of internal quantum incompatibility of states in contrast to
coherent information, which principally cannot be reduced to classical
representations. Nevertheless, the determination of coherent
information is genetically related to compatible information because it
is based on distinguishing a pair of compatible systems similar to the
input and output systems involved in the analysis of compatible
information. One of them is the reference system, and the second is the
input or output. Thus, the presence of mutually compatible sets of
quantum states is necessary for quantifying information of each of the
above types. However, when interacting systems are considered,
different types of information exhibit qualitatively different types of
behavior. The reason for this is that, unlike coherent information, the
presence of compatible information may be due to both purely quantum
and classical input-output correlations. Particularly, in the Dicke
problem, this circumstance results in possible existence of compatible
information (in contrast to coherent information) after the short-lived
collective Dicke state decays.

Selection of quantum states is shown to be principal for obtaining a
useful compatible information content. It is found that nonselected
information is equivalent to completely selected information which
averaged over all possible orientations the orthogonal bases of input
and output complete quantum measurements, realizing complete selection
of quantum states.

It is shown that mutual and internal compatibility, i.e., the
property of input and output quantum information being classical,
is a natural limitation of the physical content of the information
flow in an experimental setup. This circumstance enables one to
introduce a sufficiently general unified mathematical structure
corresponding to a chosen scheme of an actual physical experiment
and to quantify its information efficiency. In this situation, the
information exchange between subsystems preparing quantum
information and a measuring device is described by the
probabilistic correspondence between classical variables
determining the physical parameters of a quantum system under
study and measured output variables. Quantum information
generation and readout are represented in the general mathematical
form with two PSMs. This mathematical representation of quantum
information exchange realized experimentally looks promising for
applying the quantum information theory to physical experiments.
The approaches proposed in this paper additionally justify the
general statement on the physical concept of quantum information
mentioned in the paper title---``Quantum information is physical
too\ldots'' [50].

\section*{Acknowledgments} This work was supported in part by the Russian
Foundation for Basic Research (project no. 01-02-16311), the {\em
Fundamental'naya metrologiya} and {\em Nanotkhnologiya} State science
and technology programs of the Russian Federation, and INTAS (grant no.
00-479).


\begin{thebibliography}{99}
\bibitem{1}Gisin, N., Ribordy, G., Tittel, W., and Zbinden, H., LANL e-print
quant-ph/0101098.
\bibitem{2}Preskill, J., Lecture Notes on Physics
229: Quantum Information and Computation, http://www.the-ory.
caltech.edu/people/preskill/ph229/.
\bibitem{3}Steane, A., Rep. Prog. Phys., 1998, vol. 61, no. 2, p. 117.
\bibitem{4}Valiev, K.A. and Kokin, A.A.,
Kvantovye komp'yutery: nadezhdy i real'nost' (Quantum Computers: Hopes
and Reality), Izhevsk: NITs Regulyarn. Khaot. Din., 2001.
\bibitem{5}Whitaker, M.A.B., Prog. Quantum Electron., 2000, vol. 24, no. 1, p. 1.
\bibitem{6}Kadomtsev, B.B., Usp. Fiz. Nauk, 1994, vol. 164, no. 5, p. 449.
\bibitem{7}Klyshko, D.N., Usp. Fiz. Nauk, 1998, vol. 168, no. 9, p. 975.
\bibitem{8}Kilin, S.Ya., Usp. Fiz. Nauk, 1999, vol. 169, no. 5, p. 507.
\bibitem{9}Menskii, M.B., Usp. Fiz. Nauk, 2000, vol. 170, no. 6, p. 631.
\bibitem{10}Bargatin, I.V., Grishanin, B.A., and Zadkov, V.N., Usp. Fiz. Nauk,
2001, vol. 171, no. 6, p. 625.
\bibitem{11}Belokurov, V.V., Timofeevskaya, O.D., and Khrustalev,
O.A., Kvantovaya teleportatsiya---obyknovennoe chudo (Quantum
Teleportation As a Common Miracle), Izhevsk: NITs Regulyarn.
Khaot. Din., 2000.
\bibitem{12}Kadomtsev, B.B., Dinamika i informatsiya (Dynamics and
Information), Moscow: Redak. Zh. Usp. Fiz. Nauk, 2000.
\bibitem{13}The Physics of Quantum Information: Quantum Cryp-tography, Quantum
Teleportation, Quantum Computa-tion, Bouwmeester, D., Ekert, A., and
Zeilinger, A., Eds., New York: Springer, 2000.
\bibitem{14}Menskii, M.B., Kvantovye izmereniya i dekogerentsiya (Quantum
Measurements and Decoherence), Moscow: Fizmatgiz, 2001.
\bibitem{15}Korn, G. and Korn, T., Mathematical Handbook for Sci-entists and
Engineers, New York: McGraw-Hill, 1968, 2nd ed. Translated under the
title Spravochnik po matematike dlya nauchnykh rabotnikov i inzhenerov,
Moscow: Nauka, 1970.
\bibitem{16}Sudbery, A., Quantum Mechanics and the Particles of Nature. An
Outline for Mathematicians, Cambridge: Cambridge Univ. Press, 1986.
Translated under the title Kvantovaya mekhanika i fizika elementarnykh
chastits, Moscow: Mir, 1989.
\bibitem{17}Cabello, A., LANL e-print
quant-ph/0012089.
\bibitem{18}Landau, L.D. and Lifshits, E.M., Kvantovaya
mekha-nika: Nerelyativistskaya teoriya, Moscow: Nauka, 1974, 3rd ed.
Translated under the title Course of Theoretical Physics, vol. 3:
Quantum Mechanics: Non-Relativistic Theory, New York: Pergamon, 1977,
3rd ed.
\bibitem{19}Blokhintsev, D.I., Osnovy kvantovoi mekhaniki
(Funda-mentals of Quantum Mechanics), Moscow: Nauka, 1983.
\bibitem{20}Davydov, A.S., Kvantovaya mekhanika (Quantum Mechanics),
Moscow: Nauka, 1973.
\bibitem{21}Shannon, C.E., Bell Syst. Tech. J., 1948, vol.
27, p. 379; Bell Syst. Tech. J., 1948, vol. 27, p. 623.
\bibitem{22}Gallager, R.G., Information Theory and Reliable Com-munication,
New York: Wiley, 1968. Translated under the title Teoriya informatsii i
nadezhnaya svyaz', Mos-cow: Sovetskoe Radio, 1974.
\bibitem{23}Brukner, . and Zeilinger, A., LANL e-print quant-ph/ 0006087.
\bibitem{24}Hall, M.J.W., LANL e-print quant-ph/0007116.
\bibitem{25}Brukner, . and Zeilinger, A., LANL e-print quant-ph/ 0008091.
\bibitem{26}Jones, K.R.W., Phys. Rev. A, 1994, vol.
50, no. 5, p. 3682.
\bibitem{27}Caves, C.M. and Fuchs, C.M., LANL e-print
quant-ph/ 9601025.
\bibitem{28}Kholevo, A.S., Probl. Peredachi Inf., 1973, vol.
9, no. 2, p. 31.
\bibitem{29}Hall, M.J.W., Phys. Rev. A, 1997, vol. 55, no. 1,
p. 100.
\bibitem{30}Schumacher, B. and Nielsen, M.A., Phys. Rev. A, 1996, vol.
54, no. 4, p. 2629.
\bibitem{31}Lloyd, S., Phys. Rev. A, 1997, vol. 55, no. 3,
p. 1613.
\bibitem{32}Grishanin, B.A., Probl. Peredachi Inf., 2002, vol. 38, no.
1, p. 31.
\bibitem{33}Barnum, H., Schumacher, B.W., and Nielsen, M.A., Phys.
Rev. A, 1998, vol. 57, no. 6, p. 4153.
\bibitem{34}Grishanin, B.A. and Zadkov, V.N., Zh. Eksp. Teor. Fiz., 2000,
vol. 118, no. 5, p. 1048.
\bibitem{35}Grishanin, B.A. and Zadkov, V.N., Phys. Rev. A, 2000, vol. 62,
no. 3, p. 032303.
\bibitem{36}Grishanin, B.A. and Zadkov, V.N., Laser Phys., 2000,
vol. 10, no. 6, p. 1280.
\bibitem{37}Dicke, R.H., Phys. Rev., 1954, vol. 93,
no. 1, p. 9.
\bibitem{38}Stratonovich, R.L., Izv. Vyssh. Uchebn. Zaved.,
Radiofiz., 1965, vol. 8, no. 1, p. 116.
\bibitem{39}Zbinden, H., Brendel, J., Tittel, W., and Gisin, N., J.
Phys. A, 2001, vol. 34, no. 35, p. 7103.
\bibitem{40}Lindblad, G., Quantum Aspects of Optical Communica-tion:
Springer-Verlag Lecture Notes in Physics, Benja-ballah, C., Hirota, O.,
and Reynaud, S., Eds., Heidelberg: Springer, 1991, vol. 378, p. 71.
\bibitem{41}Holevo, A.S., LANL e-print quant-ph/9809022.
\bibitem{42}Grishanin, B.A., Izv. Akad. Nauk SSSR, Tekh. Kibern., 1973, vol. 11,
no. 5, p. 127.
\bibitem{43}Grishanin, B.A. and Zadkov, V.N., Laser Phys., 2001,
vol. 11, no. 12, p. 1088.
\bibitem{44}Brukner, \v{C}. and Zeilinger, A., Phys. Rev.
Lett., 1999, vol. 83, no. 17, p. 3354.
\bibitem{45}Glauber, R.J., Quantum Optics and Electronics, DeWitt, C.,
Blandin, A., and Cohen-Tannoudji, C., Eds., New York: Gordon and
Breach, 1965, p. 53.
\bibitem{46}Grishanin, B.A., Kvantovye sluchainye
protsessy (Quantum Stochastic Processes), http://comsim1.phys.msu.
su/index.html.
\bibitem{47}Schumacher, B., Complexity, Entropy and the Physics
of Information, Zurek, W.H., Ed., Redwood City: Addison-Wesley, 1990,
p. 29.
\bibitem{48}Kraus, K., States, Effects and Operations, Berlin: Springer,
1983.
\bibitem{49}Wald, A., in Pozitsionnye igry (Position Games), Moscow:
Sovetskoe Radio, 1967, p. 300.
\bibitem{50}Rudolph, T., LANL e-print quant-ph/9904037.
\bibitem{51}Schumacher, B.W., Phys. Rev. A, 1995, vol. 51,
no. 4, p. 2738.
\end{thebibliography}
\end{document}